\documentclass[number,sort&compress,twocolumn,5p]{elsarticle}
\usepackage[separate-uncertainty=true]{siunitx}
\usepackage{lineno,hyperref}
\usepackage{tikz}
\usepackage{amssymb,amsmath,mathtools}
\usepackage{caption}
\usepackage{xspace}
\usepackage{booktabs}
\usepackage{siunitx}
\usepackage{textcomp}
\usepackage[percent]{overpic}
\usepackage{acro}
\usepackage[list=true,listformat=simple,margin=10pt]{subcaption}
\usepackage{hepnames}
\usepackage{relsize}
\usepackage[capitalize]{cleveref}
\usepackage{csquotes}
\usepackage[english]{babel}

\input{default_pygstyle}

\usepackage{fvextra}
\usepackage{xcolor}
\usepackage{framed}
\usepackage{listings}
\usepackage{float}

\definecolor{minted@bgcolor}{rgb}{1,1,1}
\definecolor{minted@framecolor}{rgb}{0,0,0}

\floatstyle{plain}
\newfloat{listing}{htbp}{lop}
\floatname{listing}{Listing}

\newcommand{\code}[1]
           {{\texttt{\detokenize{#1}}}{}}

\journal{Nucl. Instr. Meth. A}

\newcommand{\cnstln}{Constellation\xspace}
\newcommand{\cnstlns}{Constellations\xspace}
\newcommand\CPP{C\nolinebreak\hspace{-.05em}\raisebox{.4ex}{\relsize{-3}{\textbf{+}}}\nolinebreak\hspace{-.10em}\raisebox{.4ex}{\relsize{-3}{\textbf{+}}}\xspace}

\renewcommand{\mkbegdispquote}[2]{\itshape}

\DeclareAcronym{BOR}{
  short = BOR,
  long = begin-of-run
}
\DeclareAcronym{COTS}{
  short = COTS,
  long = commercial off-the-shelf
}
\DeclareAcronym{CI}{
  short = CI,
  long = Continuous Integration
}
\DeclareAcronym{CD}{
  short = CD,
  long = Continuous Deployment
}
\DeclareAcronym{CDTP}{
  short = CDTP,
  long = \cnstln Data Transmission Protocol
}
\DeclareAcronym{CHP}{
  short = CHP,
  long = \cnstln Heartbeat Protocol
}
\DeclareAcronym{CHIRP}{
  short = CHIRP,
  long = \cnstln Host Identification \& Reconnaissance Protocol
}
\DeclareAcronym{CMDP}{
  short = CMDP,
  long = \cnstln Monitoring Distribution Protocol
}
\DeclareAcronym{CSCP}{
  short = CSCP,
  long = \cnstln Satellite Control Protocol
}
\DeclareAcronym{DAC}{
  short = DAC,
  long = digital-to-analog converter
}
\DeclareAcronym{DAQ}{
  short = DAQ,
  long = data acquisition
}
\DeclareAcronym{DCS}{
  short = DCS,
  long = detector control system
}
\DeclareAcronym{DQM}{
  short = DQM,
  long = data quality monitoring
}
\DeclareAcronym{DUT}{
  short = DUT,
  long = device under test
}
\DeclareAcronym{EDDA}{
  short = EDDA,
  long = Exchange on \& Development of Data Acquisitions
}
\DeclareAcronym{EOR}{
  short = EOR,
  long = end-of-run
}
\DeclareAcronym{FIFO}{
  short = FIFO,
  long = first-in first-out
}
\DeclareAcronym{FPGA}{
  short = FPGA,
  long = Field Programmable Gate Array
}
\DeclareAcronym{FSM}{
  short = FSM,
  long = finite state machine
}
\DeclareAcronym{H2M}{
  short = H2M,
  long = Hybrid-to-Monolithic
}
\DeclareAcronym{HAL}{
  short = HAL,
  long = hardware abstraction layer
}
\DeclareAcronym{HV}{
  short = HV,
  long = High Voltage
}
\DeclareAcronym{IAU}{
  short = IAU,
  long = International Astronomical Union
}
\DeclareAcronym{I2C}{
  short = I$^{2}$C,
  long = Inter-Integrated Circuit
}
\DeclareAcronym{JTAG}{
  short = JTAG,
  long = Joint Test Action Group
}
\DeclareAcronym{LSB}{
  short = LSB,
  long = least significant bit
}
\DeclareAcronym{LTO}{
  short = LTO,
  long = link time optimization
}
\DeclareAcronym{LVDS}{
  short = LVDS,
  long = low voltage differential signal
}
\DeclareAcronym{MADMAX}{
  short = MADMAX,
  long = Magnetized Disk and Mirror Axion eXperiment
}
\DeclareAcronym{MAPS}{
  short = MAPS,
  long = monolithic active pixel sensor
}
\DeclareAcronym{MIP}{
  short = MIP,
  long = minimum ionizing particle
}
\DeclareAcronym{MSB}{
  short = MSB,
  long = Most Significant Bit
}
\DeclareAcronym{PCB}{
  short = PCB,
  long = printed circuit board
}
\DeclareAcronym{PLL}{
  short = PLL,
  long = phase-locked loop
}
\DeclareAcronym{RFC}{
  short = RFC,
  long = Request for Comments
}
\DeclareAcronym{SMU}{
  short = SMU,
  long = source measure unit
}
\DeclareAcronym{SHELL}{
  short = SHELL,
  long = SHielded Experimental haLL
}
\DeclareAcronym{TLU}{
  short = TLU,
  long = trigger logic unit
}
\DeclareAcronym{ZMTP}{
  short = ZMTP,
  long = ZeroMQ Message Transport Protocol
}

\begin{document}
\begin{frontmatter}
\title{\cnstln: The Autonomous Control and Data Acquisition System for Dynamic Experimental Setups}


\author[desy]{Simon~Spannagel\corref{corr}}
\ead{simon.spannagel@desy.de}
\cortext[corr]{Corresponding author.}

\author[desy]{Stephan~Lachnit}
\author[dvel]{Hanno~Perrey}

\author[cern]{Justus~Braach}
\author[lu]{Lene~Kristian~Bryngemark}
\author[uhh]{Erika~Garutti}
\author[desy]{Adrian~Herkert}
\author[desy]{Finn~King}
\author[uhh]{Christoph Krieger}
\author[desy]{David Leppla-Weber}
\author[dvel]{Linus~Ros}
\author[desy]{Sara~Ruiz~Daza}
\author[fnal]{Murtaza~Safdari}
\author[lu]{Luis~G.~Sarmiento}
\author[uhh]{Annika~Vauth}
\author[nikhef]{H\aa kan~Wennl\"of}

\affiliation[desy]{
  organization = {Deutsches Elektronen-Synchrotron DESY},
  addressline = {Notkestr. 85},
  postcode = {22607},
  postcodesep={},
  city = {Hamburg},
  country = {Germany}
}
\affiliation[dvel]{
  organization = {Prevas Test \& Measurement AB},
  addressline = {Fabriksgatan 2C},
  postcode = {222 35},
  postcodesep={},
  city = {Lund},
  country = {Sweden}
}
\affiliation[cern]{
  organization = {European Organisation for Nuclear Research (CERN)},
  city = {Geneva},
  country = {Switzerland}
}
\affiliation[lu]{
  organization = {Lund University, Department of Physics},
  addressline = {Box 118},
  postcode = {22100},
  postcodesep={},
  city = {Lund},
  country = {Sweden}
}
\affiliation[uhh]{
  organization = {Universität Hamburg, Institut für Experimentalphysik},
  addressline = {Luruper Chaussee 149},
  postcode = {22761},
  postcodesep={},
  city = {Hamburg},
  country = {Germany}
}
\affiliation[fnal]{
  organization = {Fermi National Accelerator Laboratory},
  addressline = {Wilson St \& Kirk Rd},
  city = {Batavia},
  postcode = {60510},
  country = {USA}
}
\affiliation[nikhef]{
  organization = {Nikhef},
  addressline = {Science Park 105},
  city = {Amsterdam},
  postcode = {1098XG},
  country = {The Netherlands}
}

\begin{abstract}
The operation of instruments and detectors in laboratory or beamline environments presents a complex challenge, requiring stable operation of multiple concurrent devices, often controlled by separate hardware and software solutions.
These environments frequently undergo modifications, such as the inclusion of different auxiliary devices depending on the experiment or facility, adding further complexity.
The successful management of such dynamic configurations demands a flexible and robust system capable of controlling data acquisition, monitoring experimental setups, enabling seamless reconfiguration, and integrating new devices with limited effort.

This paper presents \cnstln, a flexible and network-distributed control and data acquisition software framework tailored to laboratory and beamline environments, that addresses the limitations of existing solutions.
The framework is designed with a focus on extensibility, providing a streamlined interface for instrument integration.
It supports efficient system setup via network discovery mechanisms, promotes stability through autonomous operational features, and provides comprehensive documentation and supporting tools for operators and application developers such as controllers and logging interfaces.

At the core of the architectural design is the autonomy of the individual components, called \emph{satellites}, which can make independent decisions about their operation and communicate these decisions to other components.
This paper introduces the design principles and framework architecture of \cnstln, presents the available graphical user interfaces, shares insights from initial successful deployments, and provides an outlook on future developments and applications.
\end{abstract}


\end{frontmatter}

\tableofcontents

\section{Introduction}
\label{sec:introduction}
The coordinated operation of instruments such as detectors, motorized stages, or power supplies is crucial to modern scientific experiments in order to produce reliable and reproducible results.
The individual components require configuration, synchronization and monitoring throughout the experiment, the recorded data needs to be transmitted to central storage nodes, and dependencies between instruments have to be resolved.
These tasks are realized by control and data acquisition software frameworks.

Large research facilities and permanent experimental installations frequently develop specialized solutions based on established large-scale frameworks~\cite{epics,doocs,tango,xdaq}.
These systems are designed to scale efficiently and to provide excellent long-term stability.
However, their complexity often necessitates significant engineering effort for the integration of new instruments and for the operation of the system.

This approach is commonly unattainable for smaller experiments, such as laboratory or beamline experimental setups, test beam campaigns, or detector prototype studies.
These projects require more flexible and lightweight solutions that can be adapted to changing experimental conditions quickly and by the instrument experts themselves.
Over the years, several control and data acquisition frameworks focused on small-scale experiments have been developed~\cite{midas,eudaq,eudaq2,artdaq,daqling} and remain in use today.
They are, however, often tailored to the specific needs of individual projects or require further application-specific adaptations and frequently rely on custom communication layers.
The lack of comprehensive documentation and standardized interfaces further complicates reuse beyond the original scope of the framework.

\cnstln is a flexible, network-distributed control and data acquisition framework that provides a reliable, maintainable, and easy-to-use system that eases the implementation of new instruments.
It is built atop established open-source libraries for network communication (ZeroMQ~\cite{zmq}) and data serialization (MessagePack~\cite{msgpack}), which provide a robust foundation and significantly reduce the required development effort.
Unlike similar frameworks, \cnstln does not rely on a central control server but is designed as a decentralized network whose components operate autonomously.
The well-documented and clear interface for instrument control enables rapid integration of new devices and allows scientists to connect new instruments with minimal added effort and with the choice between an implementation in \CPP or Python.
The framework is designed with flexibility in mind and targets applications ranging from laboratory test stands up to small and mid-sized experiments with several dozens of connected instruments and large data volumes.

This paper provides an overview of the \cnstln framework and its components and is structured as follows:
\Cref{sec:design} introduces the objectives and methodology of the framework development, while \cref{sec:framework} provides an overview of the framework architecture.
The functionality of its components are detailed in \cref{sec:satellite,sec:autonomy,,sec:interfaces}, while \cref{sec:application} showcases selected first applications of the framework.
A brief overview of the extensive documentation is given in \cref{sec:docs} before the concluding remarks and outlook in \cref{sec:conclusion}.


\section{Objectives \& Design Process}
\label{sec:design}
The development of control and data acquisition systems is often initiated in response to specific and immediate needs related to instrument control and monitoring, data storage, and network communication.
Consequently, either existing solutions are adapted to these requirements or new ones are developed, often without the opportunity to establish a clear system architecture.
This approach often limits the reusability of the solution in other settings or environments.

The work leading to the development of \cnstln was centered on the idea of turning this paradigm on its head.
Rather than being driven by an urgent and narrowly defined use case, a community of interested developers from diverse backgrounds collaborated within the \emph{\ac{EDDA}} initiative~\cite{edda} to design a robust and versatile framework applicable to a wide range of future use cases.

Several concepts have contributed to the formation of the framework structure and vision, which are outlined below.

\subsection{Starting From User Needs}

A preparatory phase was aimed at identifying and engaging a potential user community and forming a common vision of the project objectives and ambitions, primarily targeting colleagues at various research institutes.
Feature requests, experiences and use cases were systematically collected in the form of \emph{user stories}, which comprised brief descriptions of typical data acquisition scenarios in laboratory-scale experiments.
An illustrative example for such a user story is the following:

\begin{displayquote}
I have a temperature-controlled detector with high-voltage biasing applied~[\ldots].
If something goes wrong with the temperature control, I want the high-voltage power supply to ramp down automatically and
in a controlled manner.
\end{displayquote}

These accounts were subsequently distilled into a set of common requirements, forming a \emph{wishlist} of desired system functionalities.
The following requirements could be extracted from the example:

\begin{itemize}
    \item An alarm system to notify of anomalies.
    \item Capability of constituents to automatically react to notifications.
    \item Possibility for constituents to transition into a safe operation mode.
\end{itemize}

Similarly, other requirements were extracted from all collected user stories and compiled as an input for the discussions during
collaboration meetings.

\subsection{The Hackathon Format}

Several \ac{EDDA} collaboration meetings have been held in the form of so-called \emph{hackathons}.
Over dedicated periods of time, several developers met to collaborate in a shared workspace.
Here, sufficient preparation was key, both in terms of shaping of a common vision for the framework, and concerning the technical groundwork such as review of available libraries for network communication.
The discussed concepts were successfully tested in-situ using toy implementations.

The initial day of each hackathon was dedicated to discussing the objectives and their implications for the system architecture.
Throughout, discussions on the principles underlying the framework reemerged, for which sufficient time was always made available.

It is envisioned to perpetuate an adapted version of the hackathon principle as \ac{EDDA} collaboration meetings to ensure a continuous development of \cnstln and to serve as a forum for its user community.

\subsection{Working with RFCs}
\label{sub:rfc}

Communication protocols are at the heart of every network-distributed software framework, and interoperability relies on precise descriptions thereof.
Therefore, a design-driven approach was adopted for the software development through the use of \ac{RFC}-style protocol documents~\cite{rfc}.
For every communication protocol proposed for the framework, an \ac{RFC} document was drafted, detailing the exact behavior of the sending and receiving nodes.
This methodology facilitated parallel and independent implementation efforts across the two supported programming languages \CPP and Python, and ensured consistency and interoperability across the system.

The \ac{RFC} documents are part of the framework documentation which will be discussed in \cref{sec:docs}.

\subsection{Continuous Integration \& Deployment}

\begin{figure*}[tbp]
\centering
\includegraphics[width=\linewidth]{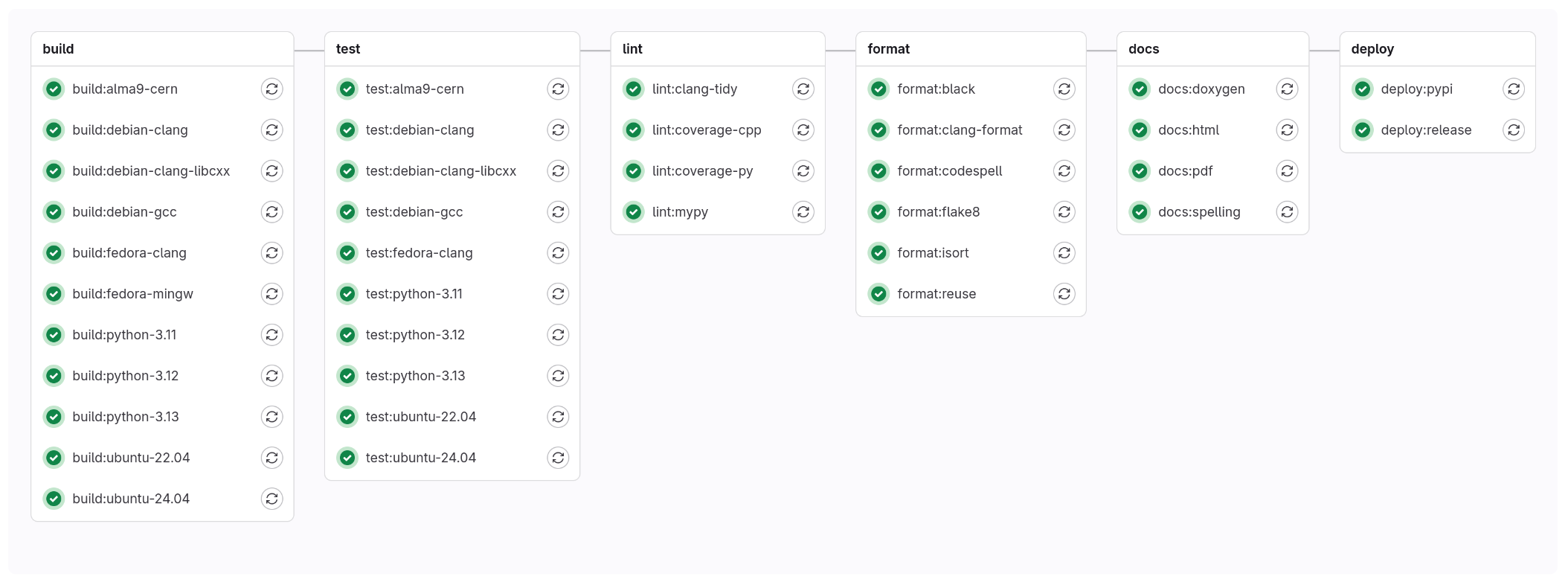}
\caption{Screenshot of the \cnstln GitLab \acf{CI} / \ac{CD} pipeline for a release version. Building, formatting and documentation generation run in parallel, while testing, linting and deployment depend on a successful build.}
\label{fig:ci}
\end{figure*}

\cnstln follows a sustainable development model which includes rigorous testing and an extensive \ac{CI} pipeline.
The release pipeline, shown in \cref{fig:ci}, commences with the \emph{build} stage which includes builds for all supported operating systems and Python versions as well as additional jobs testing different compiler and linker combinations and sanitizer libraries such as \emph{Thread Sanitizer}~\cite{tsan}.

In the \emph{testing} stage, all components of the \cnstln core libraries in both implementations are tested through unit tests with \emph{Catch2}~\cite{catch2} for \CPP and \emph{pytest}~\cite{pytest} for Python.
A combined test coverage of over \SI{85}{\percent} for both core libraries is achieved.
Static code analysis as well as formatting checks are applied in the \emph{lint} and \emph{format} stages, and the documentation is checked for spelling mistakes and built as HTML and PDF documents in the \emph{docs} stage.

The \ac{CD} comprises an automatic deployment of releases to \emph{Python Package Index PyPi}~\cite{pypi}, on \emph{Flatpak}~\cite{flatpak} and as release tarball on the GitLab repository~\cite{cnstln-repo}.
Furthermore, the website~\cite{cnstln} is updated automatically with the latest documentation upon every release.
A distribution through Homebrew for \mbox{macOS} installations is currently in preparation.


\section{Framework Architecture}
\label{sec:framework}

The name \emph{\cnstln} is borrowed from space flight, where satellite constellations operate with relatively large autonomy, transmitting data between them, and communicating with a ground control center which issues commands and controls the entire system.
Consequently, the individual components of the \cnstln control and data acquisition framework are dubbed \emph{satellites} and are complemented by \emph{controllers} and \emph{listeners} which represent user interfaces to the system.

Multiple \cnstln setups can operate in parallel on the same network without interference.
A \emph{group} name is provided to each component upon startup, and communication is only possible between components within the same group.
In this paper, the term \emph{\cnstln} either refers to the name of the software framework, or to a set of satellites operating within the same network and group.

The development repository \cnstln is hosted at the DESY GitLab instance~\cite{cnstln-repo} and a mirror is offered on GitHub~\cite{cnstln-repo-github}.
\cnstln is published under the European Union Public Licence, version 1.2 (EUPL-1.2)~\cite{eupl}.
Releases are named after \ac{IAU} designated star constellations ranked by the solid angle that they occupy in the sky.
Currently, two independent implementations of the communication protocols and the satellite code exist, written in \CPP20 and Python~3.11+.

Releases are automatically published to Zenodo~\cite{cnstln-zenodo} as software references.
This paper describes the current version \emph{\cnstln 1.0 (Corona Australis)}.

\subsection{Components of a \cnstln}
\label{sub:components}

The \cnstln framework knows three different types of components, namely \emph{satellites}, \emph{controllers} and \emph{listeners}.
Each of them has a different purpose and partakes in different communications.

\paragraph{Satellites}

These are the main constituents of a \cnstln.
They implement the code controlling an attached instrument, realizing data receivers, and any other component that should follow the \cnstln operation synchronously.
At the core of a satellite is its \ac{FSM} which governs the state of the satellite and the attached instrument.
A detailed description of the satellite is provided in \cref{sec:satellite}.

\paragraph{Controllers}

This component represents the main user interface to a \cnstln.
Controllers can send commands to satellites via the control protocol, parse and interpret configuration files, and display the current state of the entire system.
Graphical or command-line user interfaces typically are implemented as controller-type components, providing the possibility for both direct human interaction and scripted procedures.

Controllers are stateless, i.e. they are not a satellite of the \cnstln.
The main advantage of this approach is that multiple control interfaces can be active simultaneously, and that they can be closed and reopened by the operator, or even crash, without affecting the operation of the \cnstln.

The currently provided controllers are described in \cref{sub:controller} along with their features.

\paragraph{Listeners}

As the name suggests, components of this type only listen to communications of other components, typically via the monitoring protocol, and are entirely passive otherwise.
Consequently, listeners are stateless and the \cnstln is not affected by them appearing or disappearing during operations.

A typical example for a listener component is a log message interface which subscribes to logging information from satellites in the \cnstln and displays them to the operator.
Currently available listener components are presented in \cref{sub:logger,sub:telemetry}.

\subsection{System Architecture \& Protocols}
\label{sub:proto}

\cnstln is built around a set of communication protocols defining the exchange of messages among its nodes.
These protocols have been defined early on as described in \cref{sub:rfc} and serve as platform- and implementation-independent architecture of the framework.
This means that new implementations of e.g.\ a satellite can be written in any language.
The communication channels are independent of each other and follow clear communication patterns such as \emph{publish/subscribe} for one-to-many distribution of information or \emph{request/reply} for a client-server-based communication.
Most of the protocols are TCP/IP communication based on the ZeroMQ messaging library~\cite{zmq} and build upon the \ac{ZMTP}.
The messages of these protocols are encoded using MsgPack~\cite{msgpack}.

All \cnstln communication is handled exclusively via ephemeral ports as defined in RFC~6335~\cite{rfc6335}, and consequently no privileged user account is required for running a node, unless the controlled hardware requires so.
Limitations arising from this architecture will be discussed in \cref{sec:limits}.

The five \cnstln communication protocols are described in the remainder of this section.

\subsubsection{Network Discovery}

A common nuisance in volatile networking environments with devices appearing and disappearing is the discovery of available devices and services. While some established implementations based on multicast DNS~\cite{rfc6762} exist for the purpose of finding services on a local network, such as zeroconf~\cite{zeroconf} or avahi~\cite{avahi}, these come with significant downsides such as missing standard implementations, being limited to individual platforms, or a large and complex set of features not required for the purpose of \cnstln.

Other solutions, such as DIM~\cite{dim}, require central name servers to collate service information and distribute it to subscribers, a concept that conflicts with the notion of \cnstln as a decentralized network.

Hence, the \emph{\ac{CHIRP}} has been devised~\cite{cnstln-vci}.
It is an IPv4 protocol intended to be used on local networks only which uses a set of defined beacons sent as multicast messages over UDP/IP to announce or request services.
These services represent the communication endpoints of the other \cnstln communication protocols.
The beacon message contains a unique identifier for the node and its \cnstln group, the relevant service as well as IP address and port of the service. Three such beacons exist:

\begin{itemize}
    \item \code{OFFER}: A beacon of this type indicates that the sending node is offering the service at the provided endpoint.
    \item \code{REQUEST}: This beacon solicits offers for the respective service from other nodes.
    \item \code{DEPART}: A departing beacon is sent when a node ceases to offer the respective service.
\end{itemize}

\begin{figure}
\centering
\includegraphics[width=0.7\linewidth]{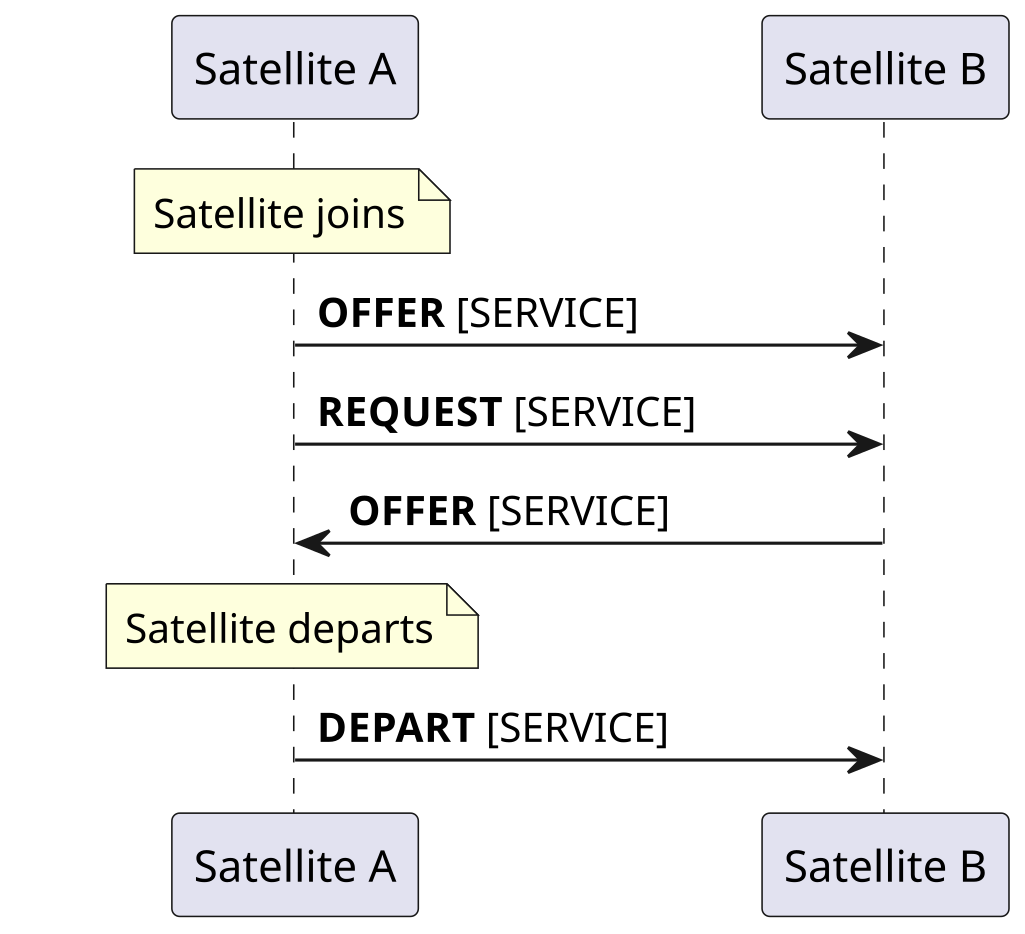}
\caption{Sequence diagram for \acf{CHIRP} showing the message flow between an already running satellite (B) and a newly started
satellite (A). Satellite A offers a service and requests a service from others. Satellite B answers the request, and finally Satellite A departs again.}
\label{fig:chirp}
\end{figure}

Each service offered by the participating \cnstln node is registered with its \ac{CHIRP} service.
Upon startup of the program, an \code{OFFER} beacon is sent for each of the registered services as indicated in \cref{fig:chirp}.
The \code{REQUEST} beacon allows nodes to join late, i.e. after the initial \code{OFFER} beacons have been distributed.
This means that at any time of the framework operation, new nodes can join and request information on a particular service from the already running \cnstln participants.
When services are shut down, the \code{DEPART} beacon will prompt other nodes to disconnect.

All subsequently described protocols use \ac{CHIRP} in order to find peers for their respective services.

\subsubsection{Heartbeat Exchange}
\label{sub:proto:chp}

Autonomous operation of the \cnstln requires the constant exchange of state information between all participants.
This is implemented in the form of heartbeat messages sent via the \emph{\ac{CHP}}.
Heartbeat messages contain the current state of the sender \ac{FSM} and the time interval after which the next heartbeat is to be expected, alongside a set of flags defining how other satellites shall react e.g.\ to changes of the \ac{FSM} state.

With this information, heartbeat receivers can independently deduce the state of other nodes, or identify non-responsive nodes in case of network or machine failure.
The variable sender-defined intervals between messages enable packet congestion control by implementing a flexible message frequency adapted to the load.
For this purpose, the sending satellite tracks the number of receivers and adjusts the interval between messages $\Delta t$ accordingly.
The interval is chosen as the minimum of $\Delta t_{\textrm{min}} = \SI{500}{\ms}$ and a configured maximum interval $\Delta t_{\textrm{max}}$ with the square root of the number of receivers $\sqrt{N}$:

\begin{equation}
\Delta t = \min \left(\Delta t_{\textrm{max}}, \max \left( \Delta t_{\textrm{min}}, \Delta t_{\textrm{min}} \cdot \sqrt{N} \cdot \ell \right) \right)\textrm{,}
\end{equation}
where $\ell$ is a load factor that allows the rate to be adjusted in the future to take additional factors into account.
It is currently set to a fixed value of~3.

The receiving node expects a new heartbeat message from each of the senders within their announced heartbeat interval.
If it fails to receive such a message in time, an internal \emph{lives} counter is decremented.
After losing three lives by not receiving a heartbeat message, the corresponding sender is considered non-responsive and appropriate action is taken.
The successful reception of a heartbeat resets this counter to its initial value.

In addition to regular heartbeat patterns, so-called \emph{extrasystoles} are sent whenever the state of the sender changes.
This enables an immediate reaction to remote state changes without having to wait for the next regular heartbeat update interval.
In case of long heartbeat intervals and a large node count, a delayed propagation of information is avoided.

\begin{figure}[tbp]
\centering
\includegraphics[width=0.7\linewidth]{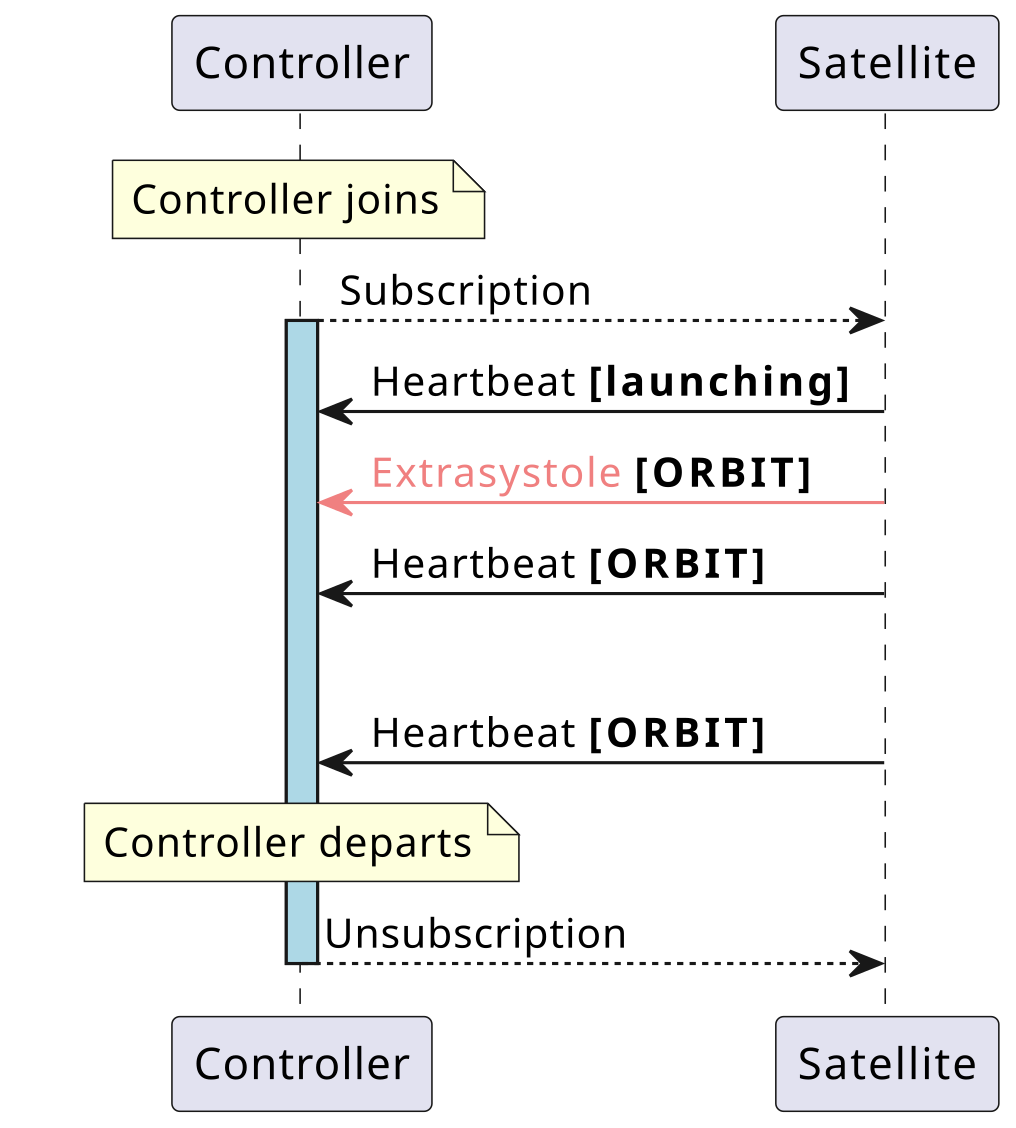}
\caption{\acf{CHP} sequence diagram between a controller and a satellite indicating regular heartbeats as well as extrasystole messages for state changes. Here, the satellite is performing a transition and issues an extrasystole once the transition is completed. The vertical colored activation bar indicates the time period of heartbeat monitoring, the bold text in square brackets denotes the transmitted state.}
\label{fig:chp:ctrl}
\end{figure}

\begin{figure}[tbp]
\centering
\includegraphics[width=0.74\linewidth]{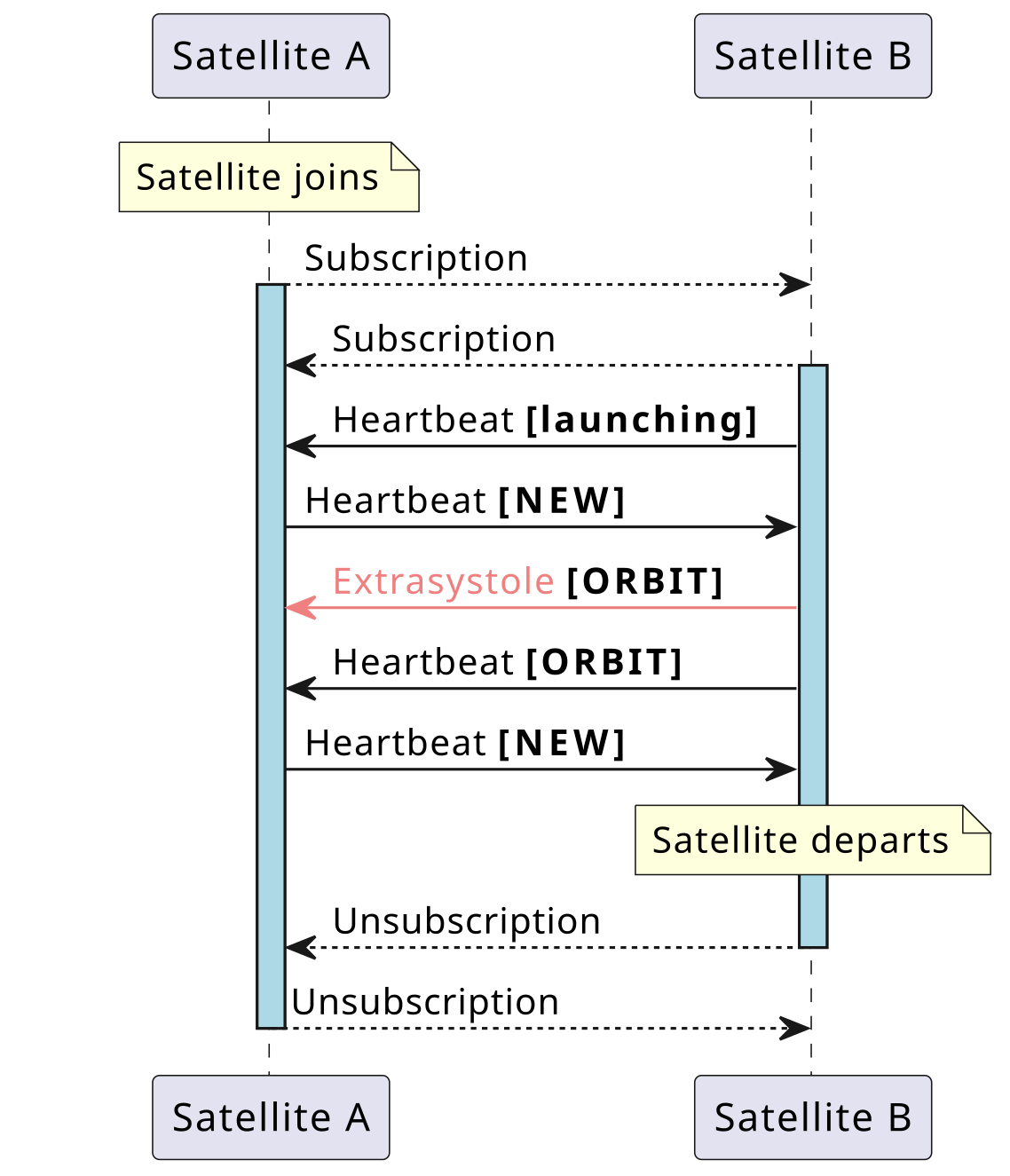}
\caption{\acf{CHP} sequence diagram for the communication between two satellites exchanging heartbeats and extrasystoles. Here, satellite B is performing a transition and emits an extrasystole once the transition is completed. The vertical colored activation bars indicate time periods of heartbeat monitoring.}
\label{fig:chp:sat}
\end{figure}

\Cref{fig:chp:ctrl} demonstrates the heartbeat exchange between a satellite and a controller instance.
The controller connects to the satellite and receives regular heartbeat messages as well as intermittent extrasystole messages for state changes.
Similarly, \cref{fig:chp:sat} shows the heartbeat exchange between two satellites with reciprocal subscription and heartbeat reception.

A detailed description of the features implemented in \cnstln based on the information transmitted by means of heartbeat messages can be found in \cref{sec:autonomy}.

\subsubsection{Control Commands}

\begin{figure}[tbp]
\centering
\includegraphics[width=0.85\linewidth]{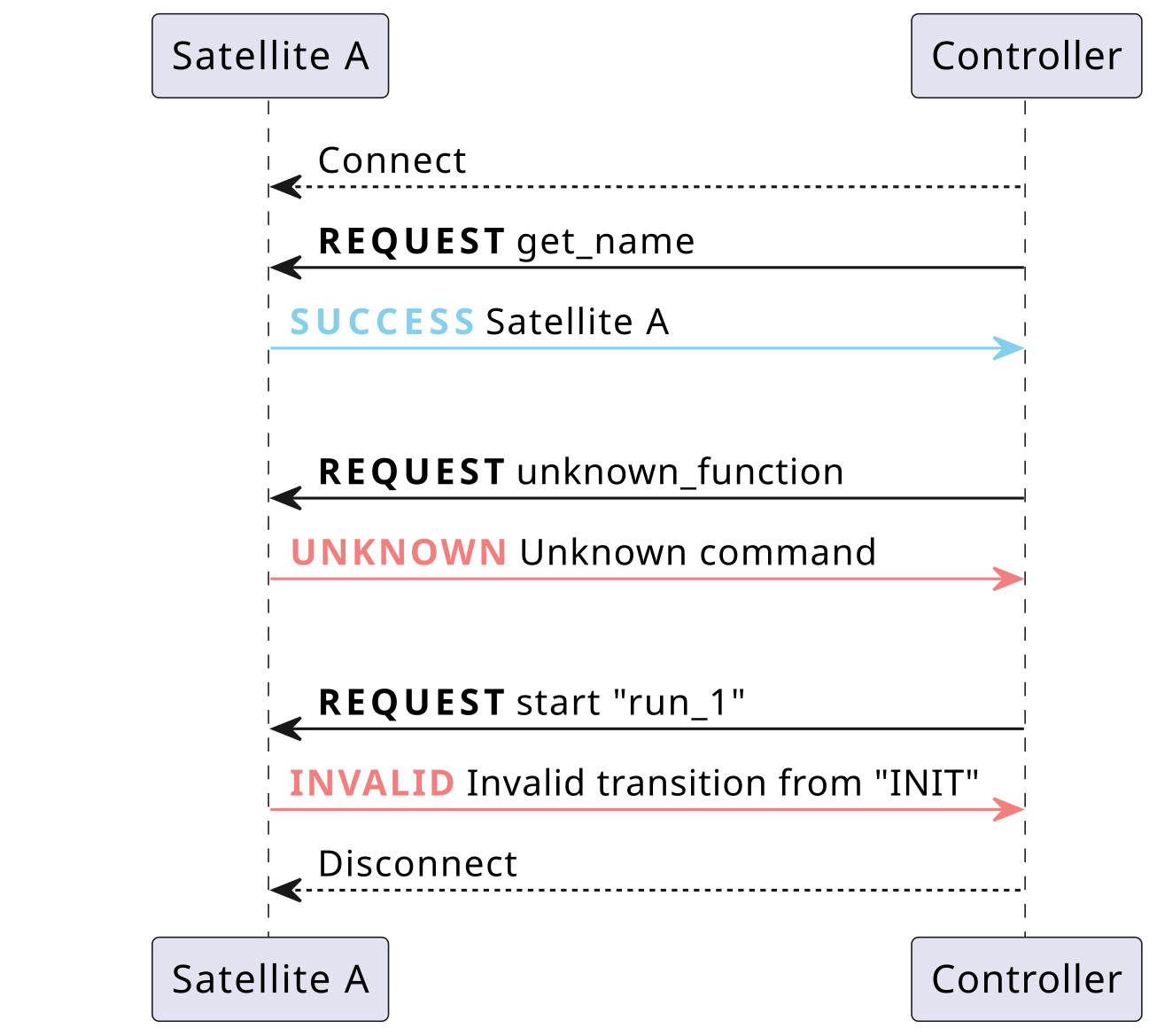}
\caption{Sequence diagram for \acf{CSCP}. A controller connects to a satellite and queries its name. Following are attempts to call an unknown function and to initiate an invalid state transition.}
\label{fig:cscp}
\end{figure}

Commands from controller instances to satellites are transmitted via the \emph{\ac{CSCP}}.
It resembles a client-server architecture with the typical request-reply pattern as indicated by the sequence diagram shown in \cref{fig:cscp}.
Here, the satellite acts as the server while the controller assumes the role of the client.
Often, controllers will also subscribe to the \ac{CHP} information of the satellites in order to directly receive real-time state updates as described above.

The command message consists of a message verb with a type and a command, and an optional payload to transmit data.
The controller exclusively sends messages with type \code{REQUEST}, while the satellite answers with a message type matching the situation, such as \code{SUCCESS} when the command was executed successfully, \code{INVALID} if for example a transition was requested that is not possible from the current state, or \code{UNKNOWN} in case the command is not known to the satellite.

The satellite command palette contains a set of standard commands to query properties and initiate state transitions, but can be extended by a specific implementation as will be described in \cref{sub:cmd}.

\subsubsection{Data Transmission}

\begin{figure}[tbp]
\centering
\includegraphics[width=0.85\linewidth]{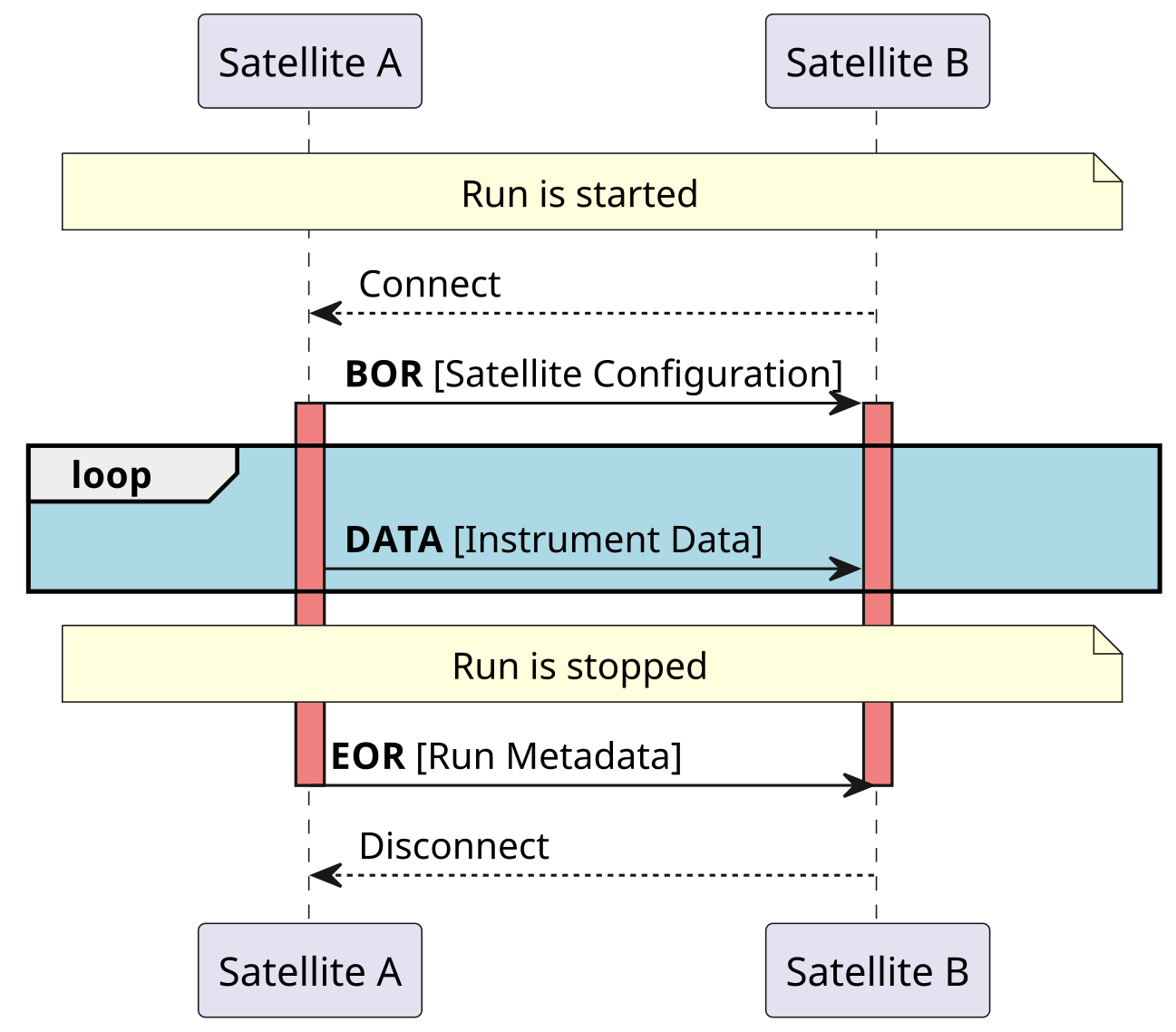}
\caption{Sequence diagram for data transmission via \acf{CDTP}. After a run is started, the receiver (Satellite B) connects to the transmitter (Satellite A). The transmitter first sends a \code{BOR} message, followed by \code{DATA} messages. Upon stopping of the run, the \code{EOR} message is sent. The colored activation bar indicates the period of active data taking.}
\label{fig:cdtp}
\end{figure}

Data are transferred within a \cnstln network using the \emph{\ac{CDTP}}.
It uses point-to-point connections via TCP/IP, as indicated in \cref{fig:cdtp}, which allow the bandwidth of the network connection to be utilized as efficiently as possible. The message format transmitted via \ac{CDTP} is a lightweight combination of a header frame with sender information and a payload that differs depending on the message type.
\ac{CDTP} knows three different message types, indicated by a flag in the header frame:

\begin{itemize}
    \item \code{BOR} - Begin of Run: This message is sent automatically at the start of a new measurement, i.e. upon entering the \code{RUN} state of the \ac{FSM} as described in \cref{sub:fsm}. It marks the start of a measurement in time and its payload frame contains the currently active satellite configuration for later reference, as well as other metadata such as the current run ID.
    \item \code{DATA}: This is the standard message type of \ac{CDTP} which consists of the header frame and any number of so-called data records containing the data of the respective instrument. Each data record may contain multiple data blocks and is marked with an incrementing data sequence counter, providing the possibility for additional offline data integrity checks.
    \item \code{EOR} - End of Run: This message is sent automatically at the end of a measurement, i.e. when the sending satellite leaves the \code{RUN} state as described in \cref{sub:fsm}. It contains metadata collected by the satellite over the course of the run.
\end{itemize}

The interface for satellites has been designed with maximal ease-of-use in mind, and does not require additional effort in satellite code to optimize data for network throughput.
Data records are passed from the satellite code to the framework library and cached in a lock-free queue~\cite{atomicqueue}.
Whenever enough data have accumulated or a configurable timeout is reached, these data records are grouped into a \ac{CDTP} \code{DATA} message and sent to their respective receivers.

\begin{figure}[tbp]
\centering
\includegraphics[width=\linewidth]{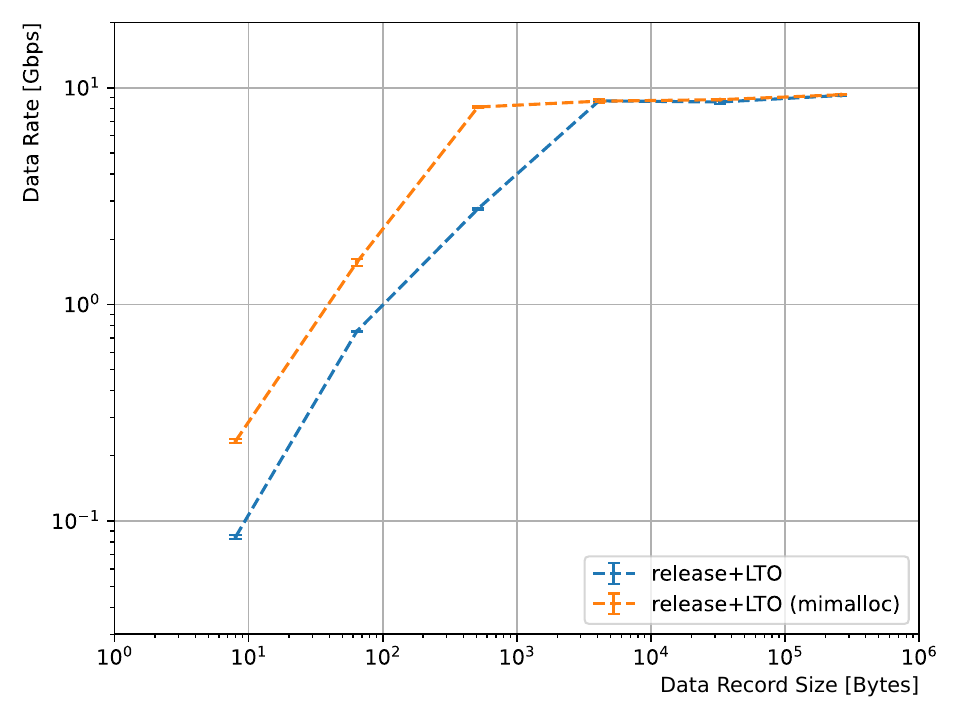}
\caption{\acf{CDTP} network throughput benchmark over a 10G fiber-optic link. Shown is the achieved data rate as a function of the data record size for two different build configurations in \emph{release} mode with \acf{LTO} activated, one using the system memory allocator and one \emph{mimalloc}~\cite{mimalloc}.}
\label{fig:cdtp-throughput}
\end{figure}

\Cref{fig:cdtp-throughput} shows a benchmark scenario run over a 10G fiber-optic network using randomly generated data.
At very small data record sizes, the data transfer is limited by memory allocation operations on the sending node as indicated by the performance difference between using the system memory allocator and an optimized allocator implementation such as \emph{mimalloc}~\cite{mimalloc}.
For record sizes above a few \si{kilobyte}, the network link bandwidth can be fully utilized.

\subsubsection{Logging \& Telemetry}

\begin{figure}[tbp]
\centering
\includegraphics[width=0.85\linewidth]{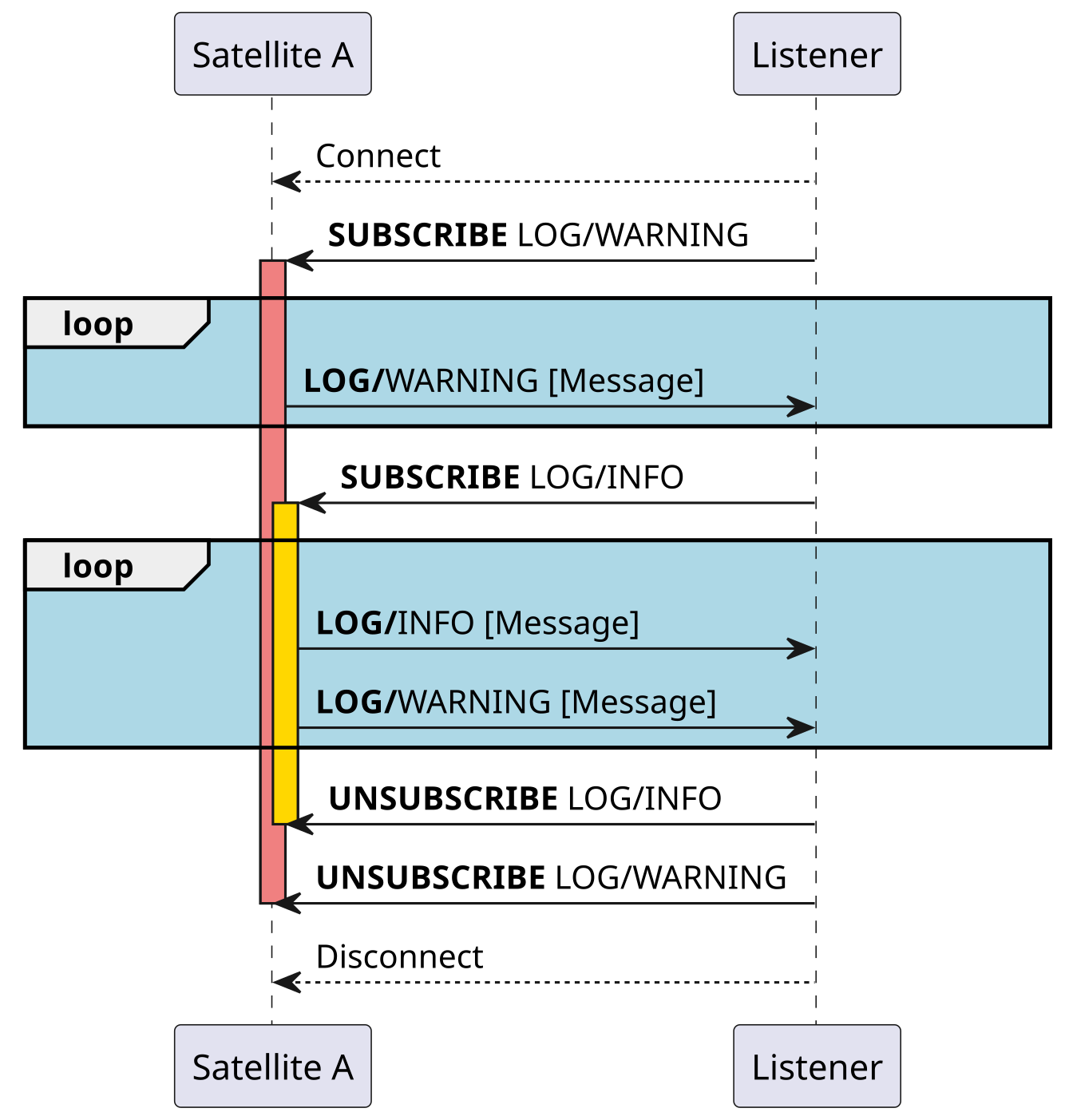}
\caption{Sequence diagram for \acf{CMDP}. A listener subscribes first to log messages of severity level \code{WARNING} and then \code{INFO} and receives messages accordingly. The colored activation bars indicate the subscription status to a certain log topic from Satellite A.}
\label{fig:cmdp}
\end{figure}

The distribution of log messages and telemetry data within \cnstln is handled by the \emph{\ac{CMDP}}.
The protocol is built around publisher and subscriber sockets which allow one-to-many distribution of messages.
The \ac{CMDP} protocol supports topics, which allows to select only the relevant slice of information from an otherwise very verbose communication, and therefore reduces the required network bandwidth.

The sequence diagram displayed in \cref{fig:cmdp} displays the transmitted information depending on the subscription.
First, a listener subscribes only to log levels with the severity \code{WARNING} and consequently only these messages are transmitted over the network.
Subsequently, the listener also subscribes to \code{INFO} messages, which increases the number of messages sent.

Subscriptions to logging levels and telemetry are completely independent of other protocols and can be performed at any time.
This means that listeners receiving log messages can be ended and restarted at will, and the subscriptions can be changed while the \cnstln is running undisturbed, enabling flexible and adaptable monitoring configurations.

The \CPP implementation of \cnstln uses \emph{spdlog}~\cite{spdlog} to efficiently queue log messages while the Python implementation relies on the native logging mechanisms.

\subsection{Limitations}
\label{sec:limits}

\cnstln targets laboratory test stands up to small and mid-sized experiments.
The focus on flexibility and ease of integrating new instruments comes with some limitations, which will be discussed in the next paragraphs.

In its current version, the framework is intended to run in closed internal networks only, in the following referred to as \emph{subnets}.
It is assumed that the subnet and all connected nodes can be trusted, that there are no malicious actors active, and that the transmitted information is non-confidential to any actor on the subset.
It is possible to configure nodes to bind to specific network interfaces only, but the default configuration---chosen for user convenience---is to bind to all available network interfaces of the node.
In general, users are required to satisfy their personal threat model by external means such as firewalls, physical isolation and virtualization on the edges of the subnet \cnstln runs on.

In order to maintain a simple interface for control as well as instrument integration, the available states and transitions of the satellite \ac{FSM} described in \cref{sub:fsm} are fixed and cannot be extended.
Furthermore, in its current state, \cnstln does not support a control hierarchy, and all satellites are directly managed by controllers.
However, future versions of the framework could implement a multi-tier control hierarchy e.g.\ by using group names to separate control domains.

Finally, the number of heartbeat message exchanges over the \ac{CHP} protocol grows quadratically with the number of nodes in the \cnstln.
While the congestion control described in \cref{sub:proto:chp} can offset performance impacts for several dozen or even hundred nodes, very large node counts may impair the response times to failure.

\section{The Satellite}
\label{sec:satellite}

The central components of a \cnstln group are satellites.
A satellite is a program responsible for controlling an instrument or providing functionality such as data storage and is built around an \ac{FSM}.
It is the only component in a \cnstln which partakes in all protocols discussed in the previous section.
In this section, the main features of the \cnstln Satellite are described.

The \emph{state} of a satellite is governed by its \ac{FSM} described in the subsequent section.
By referring to the current state of the satellite, it can be unequivocally deduced what actions the satellite is currently performing and what transitions are possible.
Sometimes, however, additional context can aid in interpreting this machine-readable state information.
This information is provided by the satellite \emph{status}, a human-readable string describing e.g. the summary of actions performed during a state transition.
In case of a failure, the status string can provide information on the cause of the error.

\subsection{Finite State Machine}
\label{sub:fsm}

The \ac{FSM} controls the behavior of the satellite and guarantees that the system is always in a defined state.
The \ac{FSM} has been designed carefully to strike the balance between robustness, features, and simplicity.
State transitions are either initiated by a controller sending transition commands through \ac{CSCP} for regular operation as described in \cref{sub:proto}, or by autonomous actions that will be discussed in \cref{sec:autonomy}.
In the subsequent paragraphs, the different states and their transitions are introduced.

\begin{figure}[tbp]
\centering
\includegraphics[width=\linewidth]{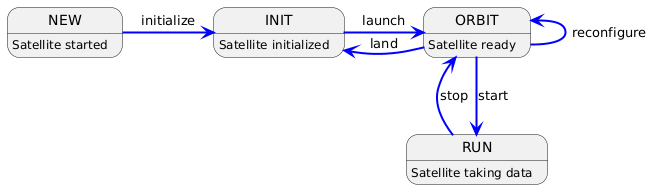}
\caption{State diagram of the satellite \ac{FSM} with the four steady states \code{NEW}, \code{INIT}, \code{ORBIT} and \code{RUN}, linked by the transition commands.}
\label{fig:fsm-steady}
\end{figure}

\paragraph{Steady States} In regular operations, i.e. without unexpected incidents in the \cnstln, the satellite \ac{FSM} will transition between four steady states.
Here, steady indicates that the satellite will remain in this state until either a transition is initiated by a controller, or a failure mode is activated.
The simplified state diagram for the normal operation mode is shown in \cref{fig:fsm-steady}.

The individual states of the satellite correspond to well-defined states of the attached instrument hardware or satellite functionality:

\begin{itemize}
    \item The \code{NEW} state is the initial state of any satellite.
          It indicates that the satellite has just been started and that no configuration has been received from a controller.
    \item The \code{INIT} state indicates that the satellite has been initialized by having received configuration data from a controller.
          A first connection to the instrument may have been made at this point.
    \item The \code{ORBIT} state signals that the satellite is ready for data taking.
          When the satellite enters this state, the instrument has been configured successfully, is powered up and is ready for entering a measurement run.
    \item The \code{RUN} state represents the data acquisition mode of the satellite.
          In this state, the instrument is active and queried for data, the satellite handles measurement data, i.e. obtains data from the instrument and sends it across the \cnstln, or receives data and stores it to file.
\end{itemize}

The \code{RUN} state implements the data taking operation of the satellite, and a \code{running} method is provided for satellite implementations to fill it with life.
The method is called upon entering the \code{RUN} state and should exit as soon as requested by the satellite \ac{FSM}.
Each run is assigned a run identifier when started, consisting of alphanumeric characters, underscores or dashes.

The anomalous steady states \code{ERROR} and \code{SAFE} will be discussed in \cref{sec:autonomy}.

\paragraph{Transitions} The code which sets up functionality or configures attached instruments is mostly executed in so-called transitional states.
Unlike steady states, they are entered by a state transition initiated through \ac{CSCP} or a failure mode, but exited automatically upon completion of the action.

\begin{figure}[tbp]
\centering
\includegraphics[width=0.9\linewidth]{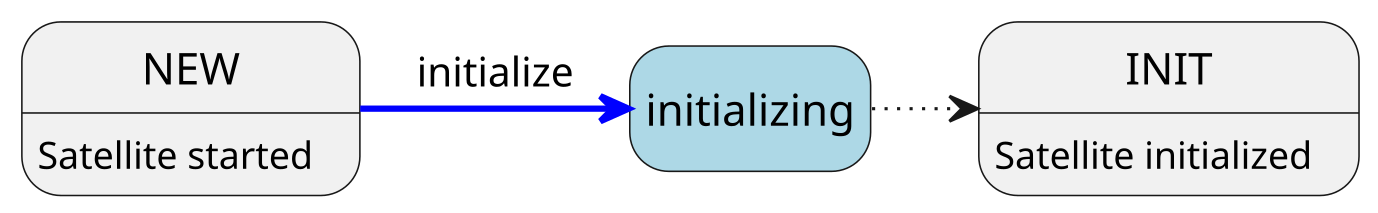}
\caption{Example for a transitional state. From the \code{INIT} state, the \code{initialize} command moves the \ac{FSM} into the \code{initializing} transitional state which automatically transits to the \code{INIT} state upon completion.}
\label{fig:fsm-transit}
\end{figure}

Such a transition diagram is shown in \cref{fig:fsm-transit}.
In this example, the transition is triggered by the \code{initialize} command sent by the controller.
The satellite enters the \code{initializing} transitional state and works through the instrument initialization code.
Upon success, the satellite automatically transitions into the \code{INIT} steady state and remains there, awaiting further transition commands.

In this scheme, actions controlling and setting up the instrument hardware directly correspond to transitional states, and each of the transition commands corresponds to entering the corresponding transitional state.
Transitional states therefore provide a direct feedback to the operator during potentially long-running actions such as the repositioning of a moveable stage or the ramping of a high-voltage power supply output.

\begin{figure}[tbp]
\centering
\includegraphics[width=0.55\linewidth]{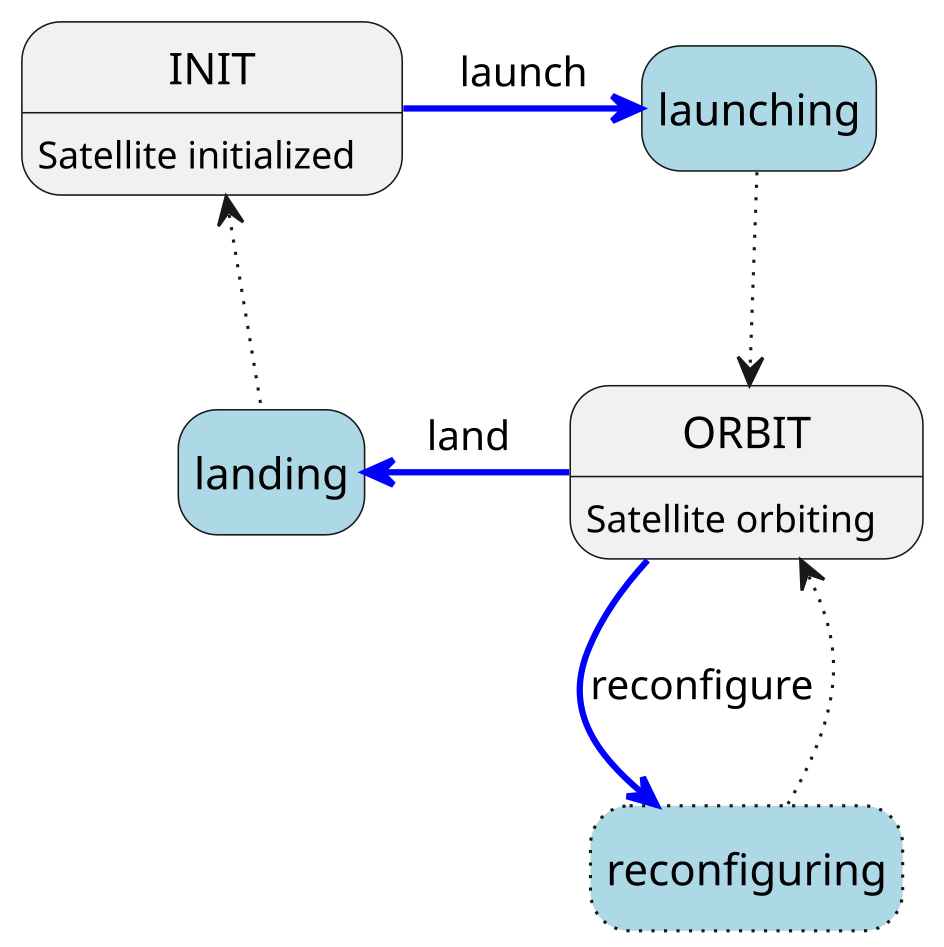}
\caption{State diagram detailing the possibilities of reconfiguring a satellite in \code{ORBIT}. Either the satellite has to go through the \code{landing}, \code{initializing} and \code{launching} states, or directly through \code{reconfiguring}.}
\label{fig:fsm-reconf}
\end{figure}

The \code{initializing} transitional states connects the steady states \code{NEW} and \code{INIT}, the states \code{launching} and \code{landing} allow switching between \code{INIT} and \code{ORBIT}, and runs are started and stopped through the \code{starting} and \code{stopping} transitional states, respectively.

The optional \code{reconfiguring} transitional state enables quick configuration updates of satellites in \code{ORBIT} state without having to pass through the \code{INIT} state as indicated in \cref{fig:fsm-reconf}.
A typical example for reconfiguration is a high-voltage power supply unit, which is slowly ramped up to its target output voltage in the \code{launching} state.
Between runs, the applied voltage is supposed to be changed by a few volts -- and instead of the time-consuming operation of ramping down via the \code{landing} transition and ramping up again, the voltage is ramped directly from its current value to the target value through the \code{reconfigure} transitional state.
This transition needs to be specifically implemented and activated in each satellite implementation in order to make it available in the \ac{FSM}.

\subsection{Controlling the Satellite}
\label{sub:cmd}

Satellites expose a set of remote procedure calls, called \emph{commands}, through the \ac{CSCP} protocol described in \cref{sub:proto}.
Controllers can query for commands using the \code{get_commands} request which returns a list of available commands and \ac{FSM} transitions along with their descriptions.
They can be divided into two types:

\paragraph{Standard commands} These commands are provided by all satellite implementations.
They comprise commands to initiate state transitions of the \ac{FSM}, e.g.\ \code{launch}, as well as methods to query additional information from the satellite such as its name via \code{get_name} or the last known run identifier via \code{get_run_id}.

\paragraph{Custom commands} Individual satellite implementations may extend this command interface by implementing custom commands specific to this satellite type and the attached instrument hardware.
Custom commands can receive input parameters through the command message payload, and can be limited to only be available in certain states of the \ac{FSM}.
A typical example for a state-limitation for a custom command is a compliance check for a high-voltage power supply.
Reading this status when the satellite is in its \code{ORBIT} or \code{RUN} state and the output voltage is turned on will produce the expected result, while reading the value in the \code{NEW} or \code{INIT} states where the power supply is not fully configured and the output voltage is off does not provide meaningful information.

\subsection{Publishing Monitoring Data}

\cnstln comes with an asynchronous monitoring mechanism which can transmit log messages of different verbosity levels as well as telemetry data over network to other nodes via \ac{CMDP} as described in \cref{sub:proto}.
In this context, asynchronous means that the code which emits a message continues immediately, while the processing of the message and its storage or transmission via network happens independently in the background.
This way, even many messages of very high verbosity level will not affect the performance of the node.
Furthermore, the subscription mechanism filters messages on the sending side, and only monitoring data with an active subscription will be transmitted.

Logging messages are distributed along with their verbosity level and a topic, in the form \code{LOG/<LEVEL>/<TOPIC>}.
The verbosity levels range from \code{TRACE} and \code{DEBUG} for detailed information on library functions to \code{WARNING} for unexpected but non-fatal problems and \code{STATUS} for high-level status information such as state changes.

Some verbosity levels contain a lot of information.
In order to allow further filtering of messages on the sending side, the topics divide the messages of each verbosity level into sections to which nodes can subscribe individually.
An example subscription to \code{LOG/TRACE/FSM} would log messages on the very verbose \code{TRACE} level, but further constrained to the \ac{FSM}, discarding all other messages.

Similarly, the telemetry is distributed with topic information following the pattern \code{STAT/<TOPIC>}.
This allows nodes to subscribe to individual metrics, e.g.\ \code{STAT/TEMPERATURE} for a temperature monitoring endpoint.

The satellite interfaces provide convenience macros which allow sending metrics in regular time intervals or only whenever the value has changed.

Notification messages are published under the subscription topics \code{LOG?} and \code{STAT?}.
They are automatically distributed whenever the available topics of the sending node have changed.
This allows receivers such as user interfaces to conveniently display all available topics to the user.

\section{Autonomous \& Collaborative Operation}
\label{sec:autonomy}

\cnstlns operate in a decentralized manner, each satellite runs autonomously and no central server which manages a common state is required.
Instead, the communication of state information between the individual components of a \cnstln is performed directly through the heartbeat messages introduced in \cref{sub:proto}.
The information available from these heartbeats can be used for a series of features introduced in the following.

\subsection{Failure Modes \& Safe State}

The autonomous satellite operation mandates a mechanism to deal with unexpected events occurring in the operation of a single satellite, but also within the entire \cnstln.
For this purpose, the satellite \ac{FSM} knows two additional steady states, also referred to as \emph{anomaly states}, reserved for anomalous situations:

The \code{ERROR} state is entered whenever an unexpected event occurs within the satellite code, the instrument control or the data transfer.
This state can only be left by a manual intervention via a controller by resetting the satellite back into its \code{INIT} state.

The \code{SAFE} state on the other hand, is entered by the satellite when detecting an issue with another satellite in the \cnstln.
This state resembles the \emph{safe mode} of an uncrewed spacecraft, where all non-essential systems are shut down and only essential functions such as communication remain active.
For a satellite this could encompass powering down instruments or switching off power supply outputs.
When no specific interrupt transition is implemented, this defaults to passing through the \code{stopping} and \code{landing} transitions described in \cref{sub:fsm}.
Also this state can only be left by a manual intervention via a controller, and the satellite will transfer back to its \code{INIT} state.

The main difference between the two failure states is the possible statement about the condition of the respective satellite.
The \code{SAFE} state is achieved via a controlled shutdown of components and is a well-defined procedure, while the \code{ERROR} state is entered, for example, through a lack of control or communication with the instrument and therefore does not allow any statement to be made about the condition of attached hardware.

\subsection{Satellite Autonomy \& Roles}

A satellite can convey information to peers by three means: by sending heartbeat messages or by ceasing to send them, by altering the state information transmitted via the heartbeat messages, and by departing from the \cnstln using a \ac{CHIRP} \code{DEPART} message.
This information allows other satellites to react to a communicated anomaly state, the absence of heartbeats for a defined period of time or a departure of the satellite as detailed in \cref{sub:proto:chp}.

In order to control the reaction of other satellites in case of irregularities, each heartbeat message contains a set of flags detailing the desired treatment of the sending satellite.
To ease the configuration of a \cnstln for operators, these flags are grouped into so-called \emph{roles} which can be assigned to each satellite individually through the configuration provided with the \code{initialize} command.
The following roles are defined:

\paragraph{\code{ESSENTIAL}}
Satellites with this role are treated most strictly by their peers.
Every reported issue will lead to an interruption of operation, including reported anomaly states, missing heartbeats or an orderly departure via \ac{CHIRP}.
If the incident appears during a run, this run will be be marked as \emph{degraded} in the \code{EOR} metadata described in \cref{sub:proto}.
A typical example would be a satellite controlling a power supply.

\paragraph{\code{DYNAMIC}}
This role is less strict with respect to orderly departure.
An incident reported via their state or missing heartbeats will still cause peer satellites to interrupt operation, an orderly departure using a \ac{CHIRP} message will be logged but does not stop operation.
If this occurs during data taking, the run will be marked as \emph{degraded}.
This is the default role the satellite assumes if not configured differently.

\paragraph{\code{TRANSIENT}}
Satellites in the transient role are considered dispensable to the core data taking operation of the \cnstln.
Consequently, neither their departure nor a reported anomalous state or missing heartbeats cause an interrupt of the ongoing operation.
Incidents are logged and ongoing measurement runs are marked as \emph{degraded}.
A typical example would be e.g. an auxiliary environmental temperature sensor.

\paragraph{\code{NONE}}
This role indicates entirely optional satellites.
The remaining \cnstln remains unaffected by incidents or departures of these satellites, and even ongoing runs will not reflect possible incidents in their metadata.

\subsection{Conditional Transitions}

Sometimes satellites require to be initialized, started or stopped in a specific order - for example, the instrument they control might depend on receiving a hardware clock from another component that is only available after initializing the corresponding satellite.
This constitutes a directed acyclic graph of dependencies between satellites for each of the transitions.

For this purpose, \cnstln provides the \code{_require_<transitional state>_after} configuration keywords, available for each of the transitions described in \cref{sub:fsm}.
The value is a list of other satellites that need to have successfully performed this transition before a given dependent satellite will perform its own transition action.
The respective satellite will receive these conditions from the controller via the configuration passed in the \code{initialize} command and evaluate them upon entering transitional states.

\begin{figure}[tbp]
\centering
\includegraphics[width=\linewidth]{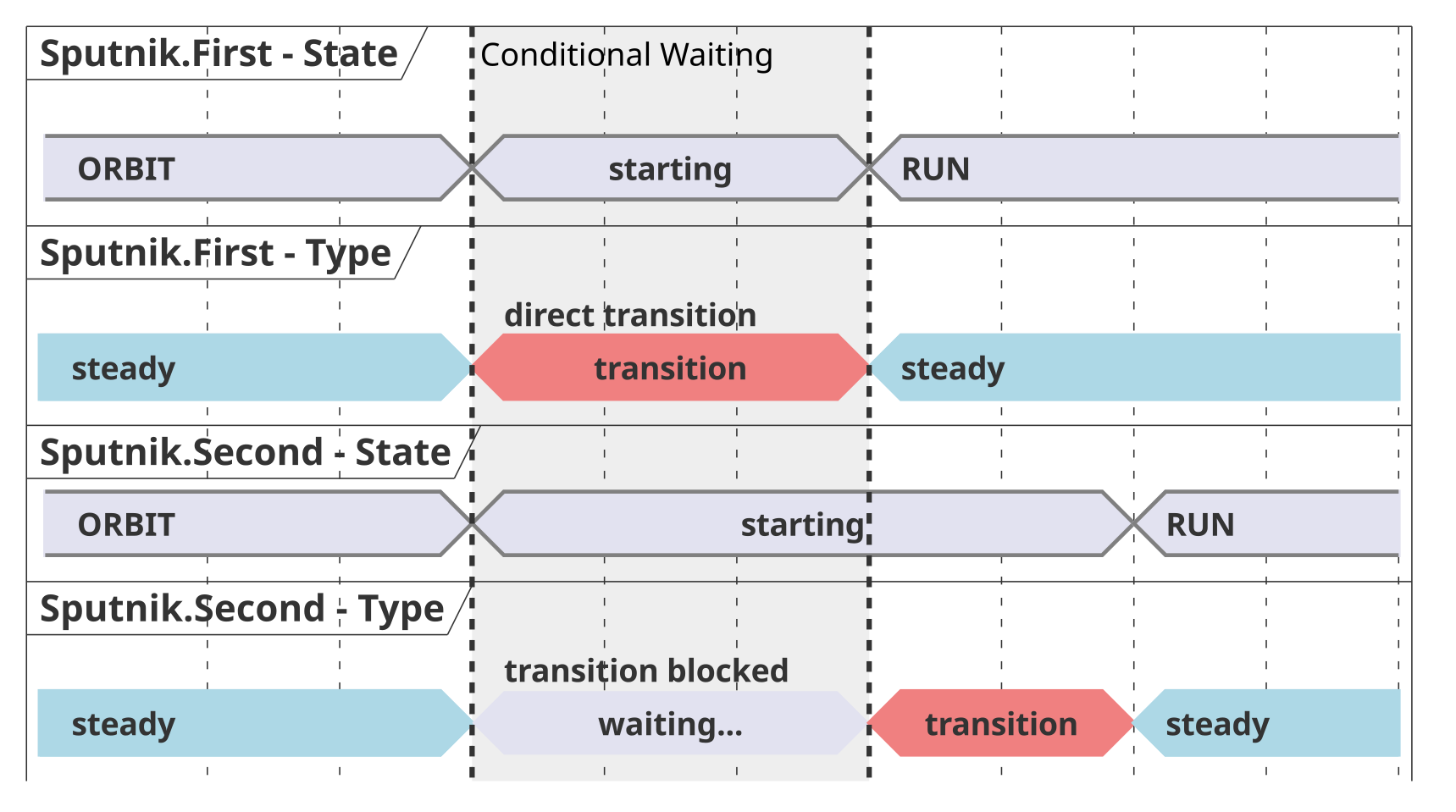}
\caption{Timing diagram of a conditional transition of satellite \code{Sputnik.Second} which waits for a successful transition of \code{Sputnik.First}.}
\label{fig:conditional-transition}
\end{figure}

If, for example, the satellite \code{Sputnik.Second} receives the condition

\begin{Verbatim}[commandchars=\\\{\}]
\PYG{n}{\PYGZus{}require\PYGZus{}starting\PYGZus{}after}\PYG{+w}{ }\PYG{o}{=}\PYG{+w}{ }\PYG{p}{[}\PYG{l+s+s2}{\PYGZdq{}Sputnik.First\PYGZdq{}}\PYG{p}{]}
\end{Verbatim}
it will enter the \code{starting} transitional state but wait before processing through its own starting action.
The latter will commence as soon as satellite \code{Sputnik.First} has successfully completed the transition and emitted the state \code{RUN} via heartbeats as visualized in \cref{fig:conditional-transition}.

This method allows satellites to asynchronously and autonomously progress from steady state to steady state without the necessity of a controller supervising the order of action.
It also allows multiple satellites waiting for the same remote condition to progress simultaneously once the condition is fulfilled.
In contrast, a transition order orchestrated by a controller necessarily would progress sequentially.

Depending on the Constellation, not necessarily all combinations are valid and will function. For example, one satellite awaiting a hardware clock signal from another satellite, which depends on the former to finalize its initialization transition, will eventually time out.
To avoid circular dependencies and the resulting deadlocks, controllers implement a directed graph validation logic using a depth-first approach to finding cyclic dependencies.

Waiting for remote conditions to be satisfied will be interrupted by the following events:

\begin{itemize}
\item The remote satellite that the condition depends on is not present or disappeared.
\item The remote satellite that the condition depends on returns an \code{ERROR} state.
\item The waiting satellite runs into the timeout for conditional transitions.
\end{itemize}

In all cases the waiting satellite aborts the pending action and transitions into its \code{ERROR} state.



\section{User Interfaces}
\label{sec:interfaces}
\cnstln comes with a set of graphical and command-line user interfaces which will be described in this section.

\subsection{Controller Interfaces}
\label{sub:controller}

Controllers are the main interfaces through which operators can set up and alter the state of a \cnstln.
Currently, the framework provides two different controllers, a graphical user interface and a scriptable interactive command-line interface.
Further interfaces, such as a stand-alone controller for a single satellite intended for laboratory setups, are under development.

\paragraph{Graphical Controller: MissionControl}

This graphical user interface is a general purpose controller for \cnstln based on Qt6~\cite{qt6}.
The main window of MissionControl shown in \cref{fig:missioncontrol} can be divided into three parts:

\begin{figure}[tbp]
  \centering
  \includegraphics[width=\columnwidth]{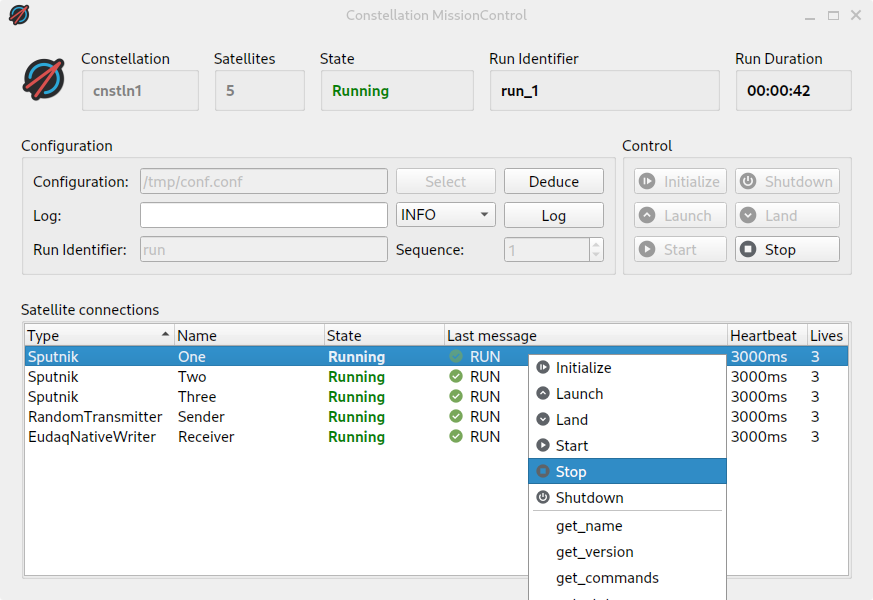}
  \caption{Main window of the MissionControl controller interface, divided into header, control area and satellite list. The header indicates that the \cnstln is in running state, and five connected nodes are shown in the satellite list, with an active context  menu for the first satellite.}
  \label{fig:missioncontrol}
\end{figure}

The header portion displays information on the \cnstln which the controller is connected to, such as its name, the number of satellites and the global state.
Here, the global state is a summarized state of all satellites.
It represents the lowest state of any individual satellite, and, if not all satellites are in the same state, is amended by the symbol~$\approxeq$ to indicate a mixed global state.

Furthermore, the current or last run identifier along with its run duration are displayed.
None of this information is stored locally; instead, it is all fetched from the running \cnstln upon startup of the controller.

The section below the header is the control area steering the entire \cnstln at once.
The input fields allow selection of the configuration file, dispatch of log messages by the operator, and setup of the run identifier.
The buttons on the right serve as control for the satellite \ac{FSM} (cf.\ \cref{sub:fsm}).
Buttons for transition commands unavailable in the current global state are deactivated and grayed out.

\begin{figure}[tbp]
  \centering
  \includegraphics[width=\columnwidth]{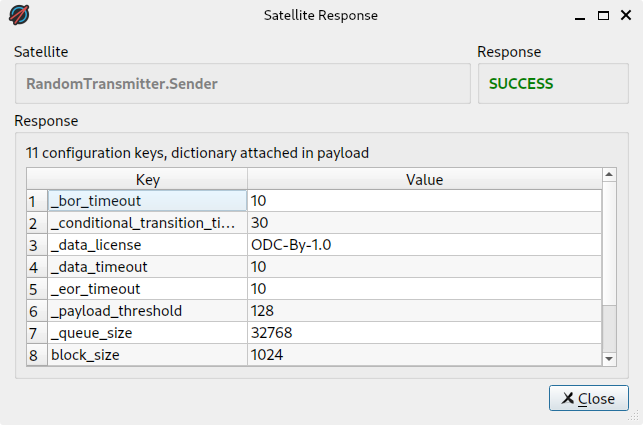}
  \caption{Satellite command response dialog of the MissionControl controller interface, displaying the command payload to the \code{get_config} command.}
  \label{fig:mcdialog}
\end{figure}

Finally, the lower part of the main window is occupied by the list of all connected satellites of the \cnstln.
The type and name of each satellite are displayed alongside its current state, the last command response and the current \ac{CHP} heartbeat interval and remaining lives.
Double-clicking a satellite item from the list opens a dialog with additional connection details.
A context menu, also shown in \cref{fig:missioncontrol}, provides direct access to the commands of the individual satellite, including transitions and possibly defined custom commands (cf.\ \cref{sub:cmd}).
All information is obtained directly from the running satellite and does not depend on any local cache or pre-loaded information in the controller.
In case a command returns payload information, such as for example the \code{get_config} command, a dialog window displays this to the operator as shown in \cref{fig:mcdialog}.

While the run identifier is a free-form string, this controller defines it as a name followed by a sequence number.
The sequence number is automatically incremented when stopping and starting a run, such that a unique run identifier is always generated and sent to the \cnstln.

Since satellites operate independently of the controller, and controllers can be started and closed at any time as described in \cref{sub:components}, they might not necessarily all have access to the same configuration file.
To alleviate this, the configuration of all satellites can be deduced from the \cnstln in operation directly via the "Deduce" button of the controller.
This will collect the current configuration from all connected satellites and open a dialog window to select a storage location for the configuration file.

\paragraph{Scriptable Controller}

In some scenarios, a scriptable command-line interface might be preferable to a graphical user interface.
For this purpose, a Python controller class is provided, both as standalone script and as interactive shell based on IPython~\cite{ipython}.
Individual satellites as well as the entire \cnstln can be queried and controlled directly through the \code{ScriptableController} class.
Complex routines such as automated parameter scans can be implemented in a few lines of Python code as demonstrated in \cref{lst:scan}.

\begin{listing}[tbp]
\setbox0=\hbox{\begin{minipage}{0.97\columnwidth}
\begin{Verbatim}[commandchars=\\\{\}, breaklines=true]
\PYG{k+kn}{import} \PYG{n+nn}{time}
\PYG{k+kn}{from} \PYG{n+nn}{constellation.core.controller} \PYG{k+kn}{import} \PYG{n}{ScriptableController}

\PYG{c+c1}{\PYGZsh{} Create controller}
\PYG{n}{ctrl} \PYG{o}{=} \PYG{n}{ScriptableController}\PYG{p}{(}\PYG{n}{group\PYGZus{}name}\PYG{p}{)}
\PYG{n}{constellation} \PYG{o}{=} \PYG{n}{ctrl}\PYG{o}{.}\PYG{n}{constellation}

\PYG{k}{for} \PYG{n}{ivl} \PYG{o+ow}{in} \PYG{p}{[}\PYG{l+m+mi}{8}\PYG{p}{,} \PYG{l+m+mi}{64}\PYG{p}{,} \PYG{l+m+mi}{128}\PYG{p}{]:}
    \PYG{c+c1}{\PYGZsh{} Reconfigure to new parameter value}
    \PYG{n}{cfg} \PYG{o}{=} \PYG{p}{\PYGZob{}}\PYG{l+s+s2}{\PYGZdq{}interval\PYGZdq{}}\PYG{p}{:} \PYG{n}{ivl}\PYG{p}{\PYGZcb{}}
    \PYG{n}{constellation}\PYG{o}{.}\PYG{n}{Sputnik}\PYG{o}{.}\PYG{n}{reconfigure}\PYG{p}{(}\PYG{n}{cfg}\PYG{p}{)}
    \PYG{n}{ctrl}\PYG{o}{.}\PYG{n}{await\PYGZus{}state}\PYG{p}{(}\PYG{n}{SatelliteState}\PYG{o}{.}\PYG{n}{ORBIT}\PYG{p}{)}
    \PYG{c+c1}{\PYGZsh{} Repeat measurement four times:}
    \PYG{k}{for} \PYG{n}{run} \PYG{o+ow}{in} \PYG{n+nb}{range}\PYG{p}{(}\PYG{l+m+mi}{1}\PYG{p}{,} \PYG{l+m+mi}{4}\PYG{p}{):}
        \PYG{n}{constellation}\PYG{o}{.}\PYG{n}{start}\PYG{p}{(}\PYG{l+s+sa}{f}\PYG{l+s+s2}{\PYGZdq{}i}\PYG{l+s+si}{\PYGZob{}}\PYG{n}{ivl}\PYG{l+s+si}{\PYGZcb{}}\PYG{l+s+s2}{\PYGZus{}r}\PYG{l+s+si}{\PYGZob{}}\PYG{n}{run}\PYG{l+s+si}{\PYGZcb{}}\PYG{l+s+s2}{\PYGZdq{}}\PYG{p}{)}
        \PYG{n}{ctrl}\PYG{o}{.}\PYG{n}{await\PYGZus{}state}\PYG{p}{(}\PYG{n}{SatelliteState}\PYG{o}{.}\PYG{n}{RUN}\PYG{p}{)}
        \PYG{c+c1}{\PYGZsh{} Run for 15 seconds}
        \PYG{n}{time}\PYG{o}{.}\PYG{n}{sleep}\PYG{p}{(}\PYG{l+m+mi}{15}\PYG{p}{)}
        \PYG{n}{constellation}\PYG{o}{.}\PYG{n}{stop}\PYG{p}{()}
        \PYG{n}{ctrl}\PYG{o}{.}\PYG{n}{await\PYGZus{}state}\PYG{p}{(}\PYG{n}{SatelliteState}\PYG{o}{.}\PYG{n}{ORBIT}\PYG{p}{)}
\end{Verbatim}
\end{minipage}}
\fcolorbox{minted@framecolor}{minted@bgcolor}{\box0}
\caption{Example for a Python controller script which scans over three values of a satellite parameter, taking four fifteen-second runs for each of them. For simplicity, this example assumes the \cnstln to be in the \code{ORBIT} state already.}
\label{lst:scan}
\end{listing}

\subsection{Receiving Log Messages}
\label{sub:logger}

Log listeners subscribe to log topics through \ac{CMDP}, to record log messages, and to present them to operators.
\cnstln offers different options to access log messages.

\paragraph{Graphical Logging Interface: Observatory}

This interface is a flexible logger for \cnstln based on Qt6~\cite{qt6}.
Its main window shown in \cref{fig:observatory} is structured in three parts:

\begin{figure}[tbp]
  \centering
  \includegraphics[width=\columnwidth]{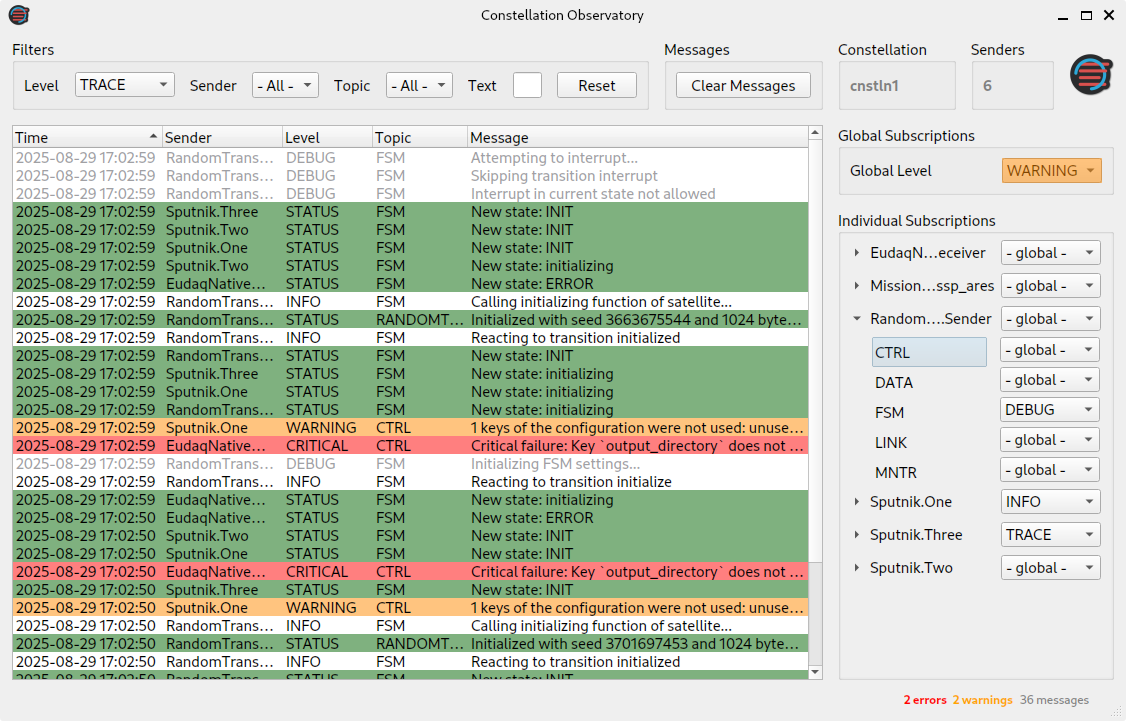}
  \caption{Main window of the Observatory logging interface showing the log messages from an attempt to initialize the \cnstln.}
  \label{fig:observatory}
\end{figure}

The central area of the window is occupied by the log message display.
Each received message is displayed as one line in the display, with its timestamp, sender name, log level, topic and message text, colored according to the log level severity.
By default, log messages appear time-ordered with the most recent message at the top.
Double-clicking a log message opens a dialog with additional information such as message tags or source code location information.

The right sidebar displays information on the \cnstln along with the number of connected log senders at the top, followed by
a drop-down to select the global log subscription level.
Below, a list of individual log senders, such as satellites or controllers, can be found, each with their own drop-down selection for log levels.
When expanding the individual senders by clicking on their name, the sender-specific list of provided \ac{CMDP} log topics is shown.
These settings provide fine-grained control over the subscribed log topics and determine which log messages are transmitted by the respective senders and received by the listener in addition to those subscribed via the global log level.
For user convenience, the log levels function as thresholds, i.e.\ any subscription automatically includes all higher-severity levels.

The filter settings for received messages are located at the top of the window.
With their help, the displayed list of log messages can be filtered by any combination of log level, sender name, log topic, or text matching the log message. It should be noted that filtering messages does not prevent them from being transmitted over the network. For this, the corresponding subscription has to be adjusted.

The status bar of the application lists the total number of received messages and indicates the number of messages received with the levels \code{WARNING} or \code{CRITICAL}.

\paragraph{Storing Log Messages: FlightRecorder}

In many application scenarios, keeping log messages for later inspection is an important feature.
\cnstln provides the \emph{FlightRecorder} satellite that implements this functionality and provides the possibility to store log messages either to a single log file or to a set of rotating log files based on their size.
Alternatively, the satellite can start a new log file every \SI{24}{\hour} at a configured time, or switch to a new log file whenever a new run is started, i.e. when it received the \code{start} command.
The latter can be especially helpful when many runs are recorded and an easy assignment of logs is required.

\paragraph{Passing Messages to Mattermost}

\begin{figure}[tbp]
  \centering
  \frame{\includegraphics[width=\columnwidth]{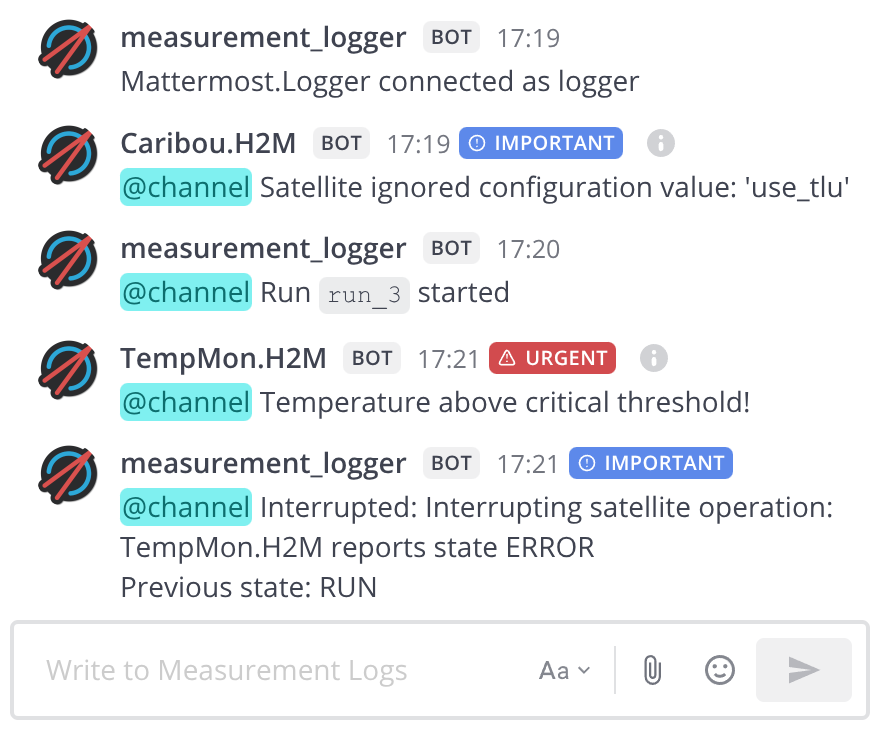}}
  \caption{Screenshot of \emph{Mattermost}, displaying a channel with log messages from \cnstln. Messages with level \code{WARNING} are marked as important, those with level \code{CRITICAL} as urgent, and are prefixed with \code{@channel} for notifications.}
  \label{fig:mattermost}
\end{figure}

Remote monitoring of the system is paramount at all times.
Therefore, \cnstln comes with a satellite which connects the \ac{CMDP} logging system with \emph{Mattermost}~\cite{mattermost}.
The satellite listens to log messages sent by other satellites and forwards them to a configured Mattermost channel.
The name of the respective satellite will be used as username to allow distinguishing them easily in the chat history.
Log messages with a level of \code{WARNING} are marked as important, messages with level \code{CRITICAL} as urgent, and both are
prefixed with \code{@channel} to notify all users in the Mattermost channel as demonstrated in \cref{fig:mattermost}.

\subsection{Monitoring of Telemetry Data}
\label{sub:telemetry}

Alongside log messages, also the monitoring of telemetry data is of vital importance to gauge the state of the \cnstln.
Telemetry listeners allow the operator to subscribe to metric topics emitted by the satellites, and to graphically display them  in a variety of charts.
Currently, there are two interfaces which allow graphical representation of metrics.

\paragraph{Real-time Monitoring: Telemetry Console}

\begin{figure}[tbp]
  \centering
  \includegraphics[width=\columnwidth]{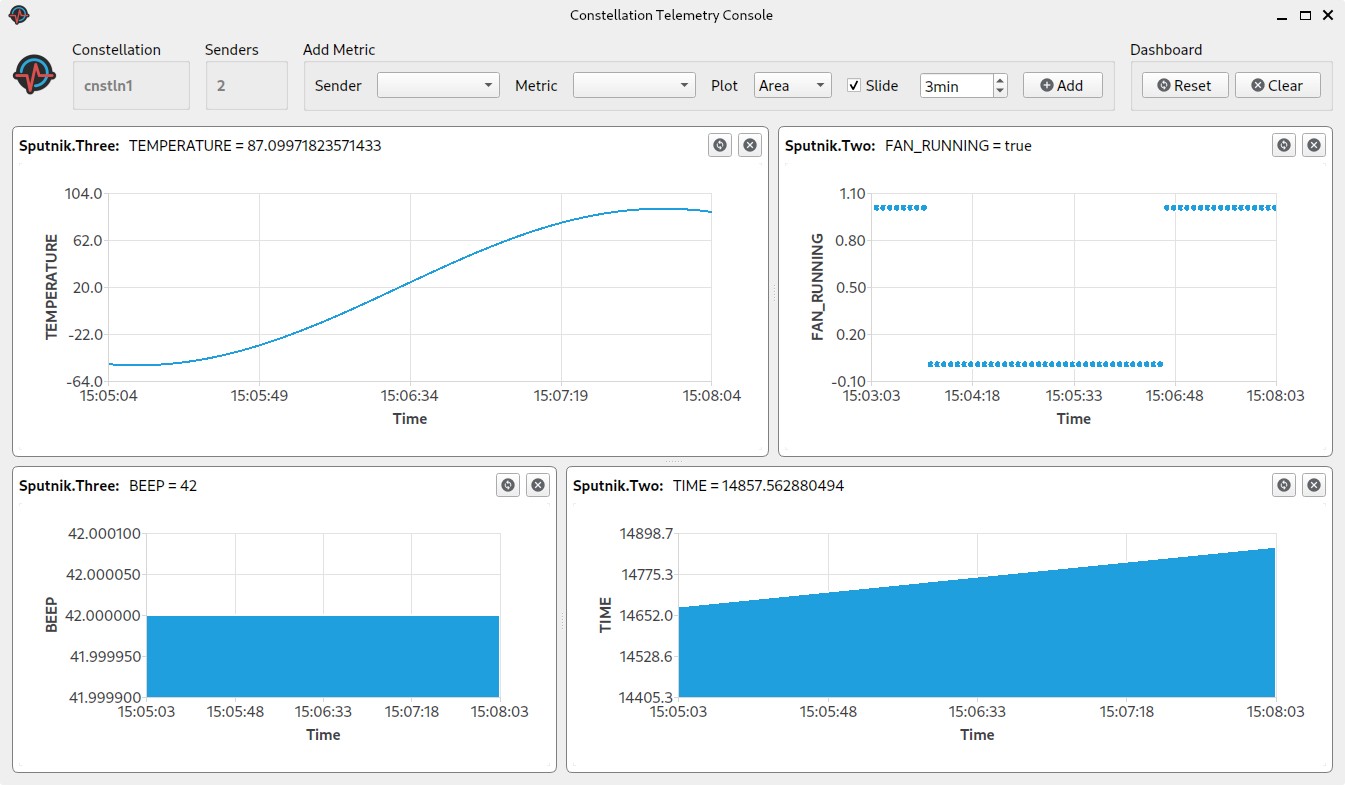}
  \caption{Main window of the Telemetry Console, displaying a dashboard with different types of graphs for real-time satellite metrics.}
  \label{fig:telemetryconsole}
\end{figure}

This graphical user interface provides a configurable dashboard to display multiple metric time series in real time.
Using the toolbar at the top of the window shown in \cref{fig:telemetryconsole}, operators can add new graphs to the dashboard area below.
The interface offers different types of graphs such as spline, scatter plot or area, and the individual graphs can be resized by dragging the dividers with the mouse.
Upon closing of the window, the current state of the dashboard -- including the displayed graphs, connected nodes and the dashboard layout -- is stored and can be restored when reopening the program.

The metrics are displayed as a time series, starting from the moment the graph is added to the dashboard.
No storage or memory of previously recorded metric information is implemented.

\paragraph{Archiving Telemetry Data: InfluxDB \& Grafana}

\begin{figure}[tbp]
  \centering
  \includegraphics[width=\columnwidth]{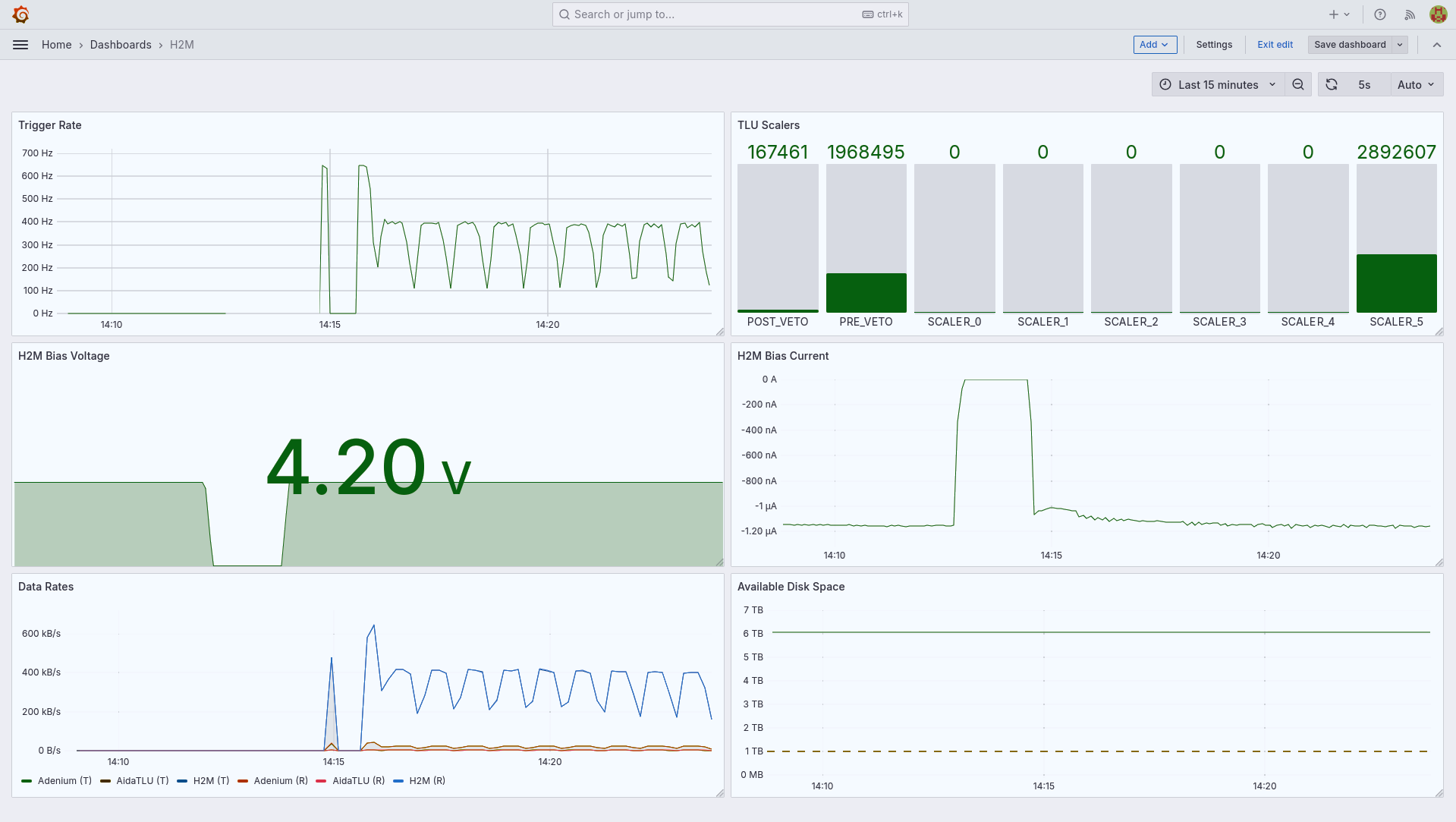}
  \caption{Example of a Grafana Dashboard displaying telemetry data of a \cnstln. Shown are trigger rate information emitted by a satellite controlling the AIDA-2020~TLU~\cite{tlu}, voltage and current information from a Keithley~2410 power supply satellite, and data rate and disk space information from a \cnstln receiver satellite.}
  \label{fig:grafana}
\end{figure}

For long-term archiving of telemetry data, the \emph{Influx} satellite, provided as part of \cnstln, can be used paired with an \emph{InfluxDB} time series database~\cite{influxdb} and a \emph{Grafana} instance~\cite{grafana}.
The satellite listens to metrics sent by other satellites and writes them to the \emph{InfluxDB}.
\Cref{fig:grafana} shows an example of a user-configured Grafana dashboard with a variety of telemetry information from a \cnstln, such as trigger information from an AIDA-2020~\ac{TLU}~\cite{tlu}, as well as voltage and current information from a Keithley~2410 \ac{SMU} satellite.
The satellite classes of \cnstln already implement several basic metrics such as data rates or available disk space for storage.

\section{First Applications in Experimental Setups}
\label{sec:application}

In this section, selected examples of early applications of the \cnstln framework in different deployment scenarios are presented, ranging from tracking of highly energetic particles to characterization of spent nuclear fuel and monitoring of a cryostat.

\subsection{Sensor Testing at the DESY~II Testbeam}

\cnstln was used to operate a particle tracking telescope and novel detector prototype at the DESY~II Testbeam Facility~\cite{desytb}.
The prototype, the so-called \emph{\ac{H2M}} chip~\cite{h2m_proceedings}, is a \ac{MAPS} manufactured in a \SI{65}{\nano\meter} CMOS imaging process.
The goal of the measurement campaign was to study its performance in terms of efficiency and resolution as a function of different operation parameters.

The experimental setup comprises a beam telescope, a timing reference detector, a \ac{TLU} as well as power supplies.
Data for reconstructing reference trajectories of the incoming particles is provided by the \emph{ADENIUM} beam telescope~\cite{adenium}.
The timing reference consists of a \emph{Telepix2} chip~\cite{telepix2}, positioned downstream of the telescope, which also serves as a region-of-interest trigger.
The \ac{H2M} prototype, placed at the center of the beam telescope, is operated and read out by the Caribou \ac{DAQ} system~\cite{caribou}, and receives a reverse-bias from a Keithley~2410 \ac{SMU}.
Trigger signals are discriminated and synchronously distributed to all detector systems by the AIDA-2020~\ac{TLU}~\cite{tlu}.
Configuration and readout of all components are coordinated using the \cnstln framework, replacing the previously used EUDAQ2 framework~\cite{eudaq2}.

A total of nine satellites were connected to the \cnstln for data taking and monitoring.
The \ac{DAQ} system satellites for H2M, ADENIUM, Telepix2 and the \ac{TLU} were ported from EUDAQ2 producers to the \cnstln framework, and a receiver satellite was implemented to store data in a EUDAQ2-compatible file format.
A \emph{Keithley} satellite controls the bias voltage applied to the H2M sensor.
In addition, an \emph{Influx} satellite was used to store telemetry data for display in a Grafana interface, and a \emph{Mattermost} satellite to keep operators informed.
A Grafana display of this measurement campaign is shown in \cref{fig:grafana}, indicating the trigger rate following the DESY~II accelerator duty cycle and its correlation with the data rates.
The number of generated trigger signals post- and pre-veto stage of the \ac{TLU} are displayed, along with the input trigger signals from Telepix2.
The interface also shows the sensor bias voltage and leakage current, the data transmission rate and the remaining available disk space.

Stable operation of the setup was achieved over several days of the measurement campaign.
Data reconstruction and analysis were performed with the Corryvreckan framework~\cite{corry}.
Compared to data previously recorded with EUDAQ2, no changes to the analysis chain were required and consistent results were achieved, while stability, system monitoring and operation were significantly improved.

\subsection{Bent Crystals at the CERN SPS North Area}

The \emph{BIPXL} telescope is a precision tracking system developed at the CERN SPS North Area Testbeam for the characterization of bent silicon crystal assemblies~\cite{decryce}.
Bent crystals manipulate particle beams for applications such as crystal-assisted collimation for heavy-ion beam~\cite{ionbeamcollim}, shadowing of the SPS electrostatic septum for loss reduction~\cite{septumlossreduction}, and future fixed-target experiments in HL-LHC~\cite{fixedtarget}.
The characterization of crystals in particle beams verifies essential specifications before installation.
Key observables are the channeling efficiency relative to particle impact angle with respect to the crystalline plane orientation, track bending angle, and torsion of the crystal assembly.

An accurate selection of tracks inside a critical angle with respect to the crystal channeling planes requires angular precision within a few \si{\micro\radian} for both incoming and outgoing particle tracks.
To achieve such performance, the telescope consists of two arms, each spanning \SI{10}{\meter} with vacuum pipes between its planes.
The instrumentation includes up to six Timepix3 pixel detectors~\cite{timepix3} and two scintillator-PMT trigger pairs, following a signal- and data-handling approach largely similar to the DESY~II tracking setup described above.

Each reference detector plane position is adjustable via paired x/y Zaber motion stages~\cite{zaber}, for reference detector alignment and extended angular acceptance of the downstream arm.
The crystal under test is centrally positioned on a four-axes goniometer by PI~\cite{pi} that allows precise alignment of its channeling planes relative to the beam.

System integration and control rely on the \cnstln framework, with custom satellites managing the Timepix3 readout and up to 14~motion stages.
Crystal alignment routines and operator feedback utilize two Corryvreckan-based satellites~\cite{corry}.
\cnstln facilitated rapid deployment at CERN within four weeks, achieving stable, reliable operation throughout a multi-day commissioning campaign.

\subsection{Spent Nuclear Fuel Characterization}

\cnstln has been used as the framework for the data acquisition in a prototype measurement station for spent nuclear fuel characterization~\cite{skb} at the Swedish Nuclear Fuel and Waste Management Company~(SKB)~\cite{skb_homepage}.
The data acquisition system consists of 24 Red~Pitaya STEMlab 125-14 devices~\cite{red_pitaya_web_stemlab}, one Red~Pitaya SIGNALab~\cite{red_pitaya_web_signallab}, and two CAEN Power supplies~\cite{caen_homepage}: a model NDT1470 and a SY5527 crate with two A7435 boards~installed.

Dedicated satellites have been implemented in Python for each of the two Red~Pitaya models.
The satellites are deployed on the Linux OS embedded in each Red~Pitaya board and read out data are stored in a ring buffer by the custom FPGA firmware.
The Red~Pitaya boards are equipped with Red~Pitaya Click Shields to synchronize the clocks and connect additional environmental sensors via the \SI{5}{\volt} I/O connectors.
For the two different CAEN high-voltage power supplies, one combined satellite implementation was developed based on the PyCaenHV library~\cite{pycaenhv_github}.
The \emph{Influx} satellite is utilized to store monitoring information in an InfluxDB database and to visualize the data using Grafana.

Signals arriving in any of the 98~analog detector channels are collected and processed in real-time on the FPGA of the connected Red~Pitaya device.
If a pulse is selected by the trigger logic of the channel, the result consisting of timestamp and pulse amplitude is first stored in RAM before being read out by the corresponding satellite.
The data are then transferred via a network switch to the host computer connected over a 10G fiber-optic cable.
Finally, the \emph{H5DataWriter} satellite stores the data to file in the HDF5~format~\cite{hdf5}.
A dedicated graphical user interface was implemented in PyQt~6 using the Python controller class described in \cref{sub:controller} to display and control all of the 29~satellites in the \cnstln group.

In laboratory tests, the data acquisition system was able to handle continuous rates of 2 million events per second and channel.
This corresponds to roughly \SI{800}{\mega\byte\per\second} of processed data being written to disk, not counting additional meta information added by the framework and the chosen file format.

\subsection{Cryostat Monitoring at MADMAX}

The \ac{MADMAX}~\cite{madmax,madmaxtemp, madmaxtemp2} is a dark matter experiment aiming to search for axions in the mass range of \SI{40}{\micro e \volt} to \SI{400}{\micro e \volt} using the dielectric haloscope approach~\cite{haloscope}.
The experiment consists of the so-called booster, a focusing mirror, and a microwave horn antenna connected to a sensitive receiver system.
In order to achieve the required low noise temperature for the final system, all components will be placed inside a custom cryostat, called the \ac{MADMAX} Cryostat.
It is currently being commissioned in the \ac{SHELL} at University of Hamburg, from where it will later move to CERN for data taking in the MORPURGO magnet~\cite{morpurgo}.

For commissioning, and later operating, the \ac{MADMAX} Cryostat, it is crucial to monitor and log several observables of the required infrastructure such as vacuum and gas pressures, cryogenic temperatures, liquid helium level.
\cnstln is used to set up a scalable and extensible monitoring system for the studies.
The monitored quantities are acquired by dedicated satellites communicating with the different measurement devices, and the resulting observables are recorded in an InfluxDB database by the \emph{Influx} satellite and monitored via Grafana dashboards as described in \cref{sub:telemetry}.
Additionally, the \emph{Mattermost} satellite is used to send warnings when parameters leave the nominal operation range or if measurement devices malfunction.
The network discovery functionality of \cnstln allowed easy and user-friendly addition of satellites on different computers during commissioning.

For the future it is planned to use \cnstln for the steering and monitoring of the \ac{MADMAX} booster, and possibility also for the data acquisition and receiver system, which would allow implementation of the entire experiment control and data acquisition in \cnstln.

\section{Documentation}
\label{sec:docs}
Constellation comes with extensive technical documentation which is published on the project website~\cite{cnstln} and as PDF documents.
It is divided into three separate documents that cater to specific target audiences:

\paragraph{Operator Guide}

This document targets operators and people who will set up and run \cnstln, control satellites and monitor the performance of the system.
The guide is structured in four different components serving different purposes:
The \emph{Get Started} section describes the installation procedure and initial setup.
\emph{Tutorials} teach how to use \cnstln using practical examples, starting from simple situations such as starting and controlling a single satellite, and gradually moving to more complex examples and setups.
The \emph{Concepts} section provides detailed explanation of the workings of the framework and the thoughts behind its structure.
This is not the technical documentation required by developers, and it simplifies explanations where possible.
Finally, the \emph{How-To Guides} provide concise answers on how to achieve a specific goal, such as setting up a telemetry database, or sending log messages to Mattermost using the satellite described in \cref{sub:logger}.

\paragraph{Application Developer Guide}

This document is intended for those that would like to integrate new instruments and functionality through satellites, or to implement new user interfaces.
It contains detailed instructions for implementing new satellites in \CPP as well as in Python, a detailed description of features such as the custom commands introduced in \cref{sub:cmd}, and guidance on how to integrate them into satellites.
The \emph{How-To Guides} section contains instructions such as porting a EUDAQ2 producer~\cite{eudaq2} to a \cnstln satellite, or how to use \cnstln as an external dependency in other projects.

\paragraph{Framework Development Guide}

The final document targets developers and potential contributors to the central \cnstln framework components.
It contains the technical details of the implementation, alongside the protocol RFC documents mentioned in \cref{sub:rfc}, satellite implementation guidelines, naming conventions, and the reference for the application programming interfaces of the different \cnstln components.

In addition to these three documents, the website also hosts a list of satellites available from different sources.
For each satellite, a description of the functionality as well as the available commands, metrics and configuration parameters are listed alongside their default values where applicable.
The page lists satellites that are part of the main \cnstln repository as well as implementations that are hosted externally.
The latter is the preferred option for satellites that implement control of specialized hardware, while satellites with broad applicability or those that control \acl{COTS} hardware can be submitted for inclusion in the main repository.


\section{Conclusions \& Outlook}
\label{sec:conclusion}
Constellation is a flexible, decentralized, and extensible framework for control and data acquisition in dynamic experimental setups.
By putting focus on autonomy, standardized protocols, and ease of integration, it addresses several of the limitations found in existing solutions, particularly for small to mid-sized laboratory experiments.

The design is centered around independent satellite nodes, which enables operation without depending on a central control server.
Regular heartbeat messages exchanged between the nodes enable the implementation of advanced features such as autonomous reaction to error conditions or conditional transition sequences.

The framework has already demonstrated its practical applicability in different experimental environments such as charged particle tracking, spent nuclear fuel characterization and cryostat temperature monitoring.
It is capable of coordinating multiple heterogeneous components, enables autonomous operation, and provides a stable foundation for control, system monitoring and data acquisition of small to mid-size experimental setups.

Several enhancements are under consideration for future versions of \cnstln such as an access control system to avoid state changes by mistake, and hierarchical control domains through intermediary satellites.
Future developments will continue to be community-driven, e.g.\ through EDDA hackathons, and will focus on flexibility and long-term stability of the framework.


\section*{Acknowledgements}
\label{sec:acknowledgements}
This project has received funding from the Lund University and Universität Hamburg Seed Fund 2023.
Part of the presented measurement results have been performed at the Test Beam Facility at DESY Hamburg (Germany), a member of the Helmholtz Association (HGF).

\section*{CRediT authorship statement}
\label{sec::credit}

\noindent
\textbf{J.~Braach:} Investigation, Software, Writing -- Review \& Editing
\textbf{L.~K.~Bryngemark:} Conceptualisation, Methodology, Writing -- Original draft, Writing -- Review \& Editing, Project administration, Funding acquisition.
\textbf{E.~Garutti:} Resources, Project administration, Funding acquisition
\textbf{A.~Herkert:} Validation, Writing -- Review \& Editing.
\textbf{F.~King:} Validation, Software, Writing -- Review \& Editing.
\textbf{C.~Krieger:} Supervision, Validation
\textbf{D.~Leppla-Weber} Validation, Software.
\textbf{S.~Lachnit:} Conceptualisation, Investigation, Software, Visualisation, Writing -- Review \& Editing.
\textbf{H.~Perrey:} Conceptualisation, Funding acquisition, Investigation, Software, Writing -- Review \& Editing.
\textbf{L.~Ros:} Conceptualisation, Investigation, Software, Writing -- Review \& Editing.
\textbf{S.~Ruiz~Daza:} Investigation, Writing -- Original Draft.
\textbf{M.~Safdari:} Validation, Writing -- Review \& Editing.
\textbf{L.~G.~Sarmiento:} Conceptualisation, Writing -- Review \& Editing.
\textbf{S.~Spannagel:} Conceptualisation, Methodology, Software, Investigation, Resources, Writing -- Original Draft, Writing -- Review \& Editing, Visualisation, Supervision, Project administration, Funding acquisition.
\textbf{A.~Vauth:} Conceptualisation, Project administration, Funding acquisition.
\textbf{H.~Wennl\"of:} Conceptualisation, Investigation, Software


\bibliography{bibliography}

\end{document}